\documentclass[journal]{IEEEtran}
\ifCLASSINFOpdf
\else
\fi

\usepackage{xcolor}

\usepackage{framed}

\colorlet{shadecolor}{yellow}

\usepackage[pdftex]{graphicx}
\graphicspath{{../pdf/}{../jpeg/}}
\DeclareGraphicsExtensions{.pdf,.jpeg,.png}

\usepackage[cmex10]{amsmath}
\usepackage{array}
\usepackage{mdwmath}
\usepackage{mdwtab}
\usepackage{eqparbox}
\usepackage{url}

\usepackage{amssymb}
\graphicspath{ {./images/} }
\usepackage{orcidlink}
\usepackage{breqn}
\usepackage{physics}

\usepackage{tikz} 

\usepackage{soul}   

\usepackage{enumitem}  

\usepackage{threeparttable}  

\usepackage{enumerate}

\newif\ifshowmods
\showmodsfalse   

\ifshowmods
                \newcommand{\Ared}[1]{\textcolor{black}{#1}} 
                \newcommand{\Ablue}[1]{\textcolor{blue}{#1}}

\else
                \newcommand{\Ared}[1]{#1}
                \newcommand{\Ablue}[1]{#1}

\fi

\begin{document}

\title{

Unmasking Covert Intrusions: Detection of Fault-Masking Cyberattacks on Differential Protection Systems

}

\author{Ahmad~Mohammad~Saber\orcidlink{0000-0003-3115-2384},~\IEEEmembership{Member,~IEEE,}
        Amr~Youssef\orcidlink{0000-0002-4284-8646},~\IEEEmembership{Senior Member,~IEEE,}
        Davor~Svetinovic\orcidlink{0000-0002-3020-9556},~\IEEEmembership{Senior Member,~IEEE,}
        Hatem~Zeineldin\orcidlink{0000-0003-1500-1260},~\IEEEmembership{Senior Member,~IEEE,}
        and~Ehab~F.~El-Saadany\orcidlink{0000-0003-0172-0686},~\IEEEmembership{Fellow,~IEEE}
\thanks{This work was supported by Khalifa University, UAE, and by VRI20-07- thrust 4.3, ASPIRE Virtual Research Institute Program, Advanced Technology Research Council, UAE.}
\thanks{Ahmad Mohammad Saber is with the
 Department of Electrical Engineering, College of Engineering and Physical Sciences Khalifa University, Abu Dhabi, UAE
(e-mail: \href{mailto:ahmad.m.saber@ieee.org}{ahmad.m.saber@ieee.org}).}
\thanks{Amr Youssef is with the Concordia Institute for Information Systems Engineering (CIISE), Concordia University, Montreal, QC, Canada
(e-mail: 
\href{mailto:youssef@ciise.concordia.ca}{youssef@ciise.concordia.ca}).}
\thanks{Davor Svetinovic is with the Department of Computer Science, College of Computing and Mathematical Sciences, Khalifa University, Abu Dhabi, UAE (e-mail: \href{mailto:davor.svetinovic@ku.ac.ae}{davor.svetinovic@ku.ac.ae}).}
\thanks{\textcolor{black}{Hatem H. Zeineldin is with the
 Department of Electrical Engineering, College of Engineering and Physical Sciences Khalifa University, Abu Dhabi, UAE, and the Electric Power Engineering
Department, Cairo University, Giza, Egypt (email:  \href{mailto:hatem.zeineldin@ku.ac.ae}{ hatem.zeineldin@ku.ac.ae}).}}
\thanks{ Ehab F. El-Saadany is with the  Department of Electrical Engineering, College of Engineering and Physical Sciences Khalifa University, Abu Dhabi, UAE, and is an Adjunct Professor with the University of Waterloo, ON, Canada
(email: \href{mailto:ehab.elsadaany@ku.ac.ae}{ehab.elsadaany@ku.ac.ae}).}
}

\markboth{
}%
{Shell \MakeLowercase{\emph{et al.}}: Bare Demo of IEEEtran.cls for IEEE Journals}

\maketitle

\begin{abstract}
Line Current Differential Relays (LCDRs) are high-speed relays progressively used to protect critical transmission lines. However, LCDRs are vulnerable to cyberattacks. Fault-Masking Attacks (FMAs) are stealthy cyberattacks performed by manipulating the remote measurements of the targeted LCDR to disguise faults on the protected line. Hence, they remain undetected by this LCDR. In this paper, we propose a two-module framework to detect FMAs. The first module is a Mismatch Index (MI) developed from the protected transmission line's equivalent physical model. The  MI is triggered only if there is a significant mismatch in the LCDR’s local and remote measurements while the LCDR itself is untriggered, which indicates an FMA. After the MI is triggered, the second module, a neural network-based classifier, promptly confirms that the triggering event is a physical fault that lies on the line protected by the LCDR before declaring the occurrence of an FMA. The proposed framework is tested using the IEEE 39-bus benchmark system. Our simulation results confirm that the proposed framework can accurately detect FMAs on LCDRs and is not affected by normal system disturbances, variations, or measurement noise. Our experimental results using OPAL-RT's real-time simulator confirm the proposed solution's real-time performance capability. 
\end{abstract}
\IEEEpeerreviewmaketitle
\section{Introduction}
\IEEEPARstart{S}{mart} grids employ advanced information and communication technologies for various monitoring, controlling, and protective functions. However, with the popularity of these technologies, cyberattacks on smart grid components have become inevitable. For instance, malicious entities carried out cyberattacks on the Ukrainian power grid in 2015 and 2016, gaining control of the substations' protection systems and causing multiple power outages.
Digital protective relays have recently been identified as possible targets of cyberattacks \cite{DOHS}. These relays detect faults in power systems,  which must be promptly cleared to maintain system stability.
MITRE Corporation\footnote{URL: https://www.mitre.org/} has emphasized that malevolent entities may intentionally target protection devices to disable their protective functions, similar to how the Industroyer malware targeted and deactivated protection relays in 2016 \cite{MITRE, MITRE2}. 
In line with this, the U.S. National Institute of Standards and Technology provides an example of cyberattacks targeting smart-grid relays:  ``an adversary tampering with the integrity of protective relay settings before a physical attack on power lines. Although the original settings were designed to contain the effects of a failure, the tampered settings allow the failure to cascade into impacts on a wider segment of the grid'' \cite{NISTGuidelines}. The described example is a cyberattack aiming to mask faults and compromise the dependability of digital relays.

\subsection{Line Current Differential Relays and Cyberattacks}
For critical transmission lines, LCDRs are increasingly employed as primary protection due to their excellent speed,  sensitivity (i.e., how accurately they detect faults on the line of interest), protective security, and selectivity (i.e., robustness to out-of-line system disturbances) \cite{Ahmadpaper}. 
In comparison with other relays, e.g., directional overcurrent and distance relays,
LCDRs have better selectivity and higher sensitivity to high-impedance faults. In addition, LCDRs are more economical than Traveling-Wave-Based Relays (TWBRs) and do not encounter the issues associated with detecting the second traveling wave, unlike TWBRs. 
In reality, two LCDRs are installed for each line, with one near each terminal. These LCDRs exchange time-synchronized current measurements to detect internal faults on the protected line. Remote measurements, timestamped by each LCDR based on GPS signals, are exchanged over two-way communication, often wireless and/or vulnerable media like microwave and radio communication, which can expose LCDRs to cyber-induced attacks \cite{Kundur_DL}.  
Possible cyberattacks on LCDRs include the \Ared{mechanisms:} (i) Time-Synchronization Attacks (TSAs), in which the adversary spoofs the GPS signal or remotely attacks the time synchronization mechanisms employed by LCDRs \cite{Remedial_pilot_main_protection},
 (ii) Denial of Service Attacks (DoSAs), where the opponents drop the communication link between a pair of LCDRs to block their operation \cite{ResMVDC}, and 
(iii) False-Data Injection Attacks (FDIAs), in which malicious entities remotely inject falsified measurements into the communication channel carrying the remote measurements of an LCDR 
\cite{Ahmadpaper, SEpaper_original}.

An FDIA targeting an LCDR can have one of two goals. Firstly, the FDIA can aim to cause unnecessary tripping when the power system is healthy, which is known as a false-tripping attack (FTA). On the contrary, an FDIA can aim to mask an actual fault from a primary relay so the fault goes undetected by this relay. This type of FDIA is referred to in this paper as a Fault-Masking Attack (FMA).
On the one hand, FTAs impact the power system's reliability as they can trigger the relay to trip under the healthy operation of the system.
On the other hand, an FMA, when performed against a relay, aims to prevent the relay from sensing a fault on the line the attacked relay is supposed to protect. Thus, FMAs can impact the security and stability of the power system, as undetected faults can result in power system instability, damage to physical equipment, fires, and electric shocks to humans. 
FTAs are identified by distinguishing them from genuine faults \textit{after} the relay has been triggered \cite{Ahmadpaper}. However, this approach is ineffective for detecting FMAs, as they intentionally keep the relay untriggered. Therefore, there is an urgent need to develop solutions dedicated to detecting FMAs.

\subsection{Related Works and Research Gap}

\Ablue{The cybersecurity of various power system components, such as Load Frequency Control schemes and power system state estimation, has been extensively studied, with a focus on defending against cyberattacks like Denial of Service Attacks and deception attacks \cite{TSMC_LFC, TSMC_LFC2, TSMC_SE_MTD}. However, the unique and critical threat posed by FMAs on LCDRs  has not been sufficiently addressed, despite its potential to severely disrupt power system stability and safety. Existing research on fault-concealing attacks has primarily targeted Wide-Area Protection (WAP) systems \cite{Shahidehpour_Concealing,  Shahidehpour_FMA}, leaving a significant gap in the protection of LCDRs from FMAs.}
Cybersecurity of overcurrent relays was investigated in 
\cite{Pola_Azzouz_Overcurrent}. In \cite{European_patent}, communication-layer-based security measures were proposed for digital relays. Authors of \cite{CB_patent} proposed a voltages-polarity-based method to differentiate between legitimate and malicious false-tripping and/or fault-masking control signals of circuit breakers. Given the criticality of LCDRs, several recent studies have proposed solutions to be implemented within LCDRs to detect potential cyberattacks, thus enhancing their security with defense-in-depth layers. A TSA-immune remedial communication-dependant protection scheme has been proposed \cite{Remedial_pilot_main_protection}. 
DoSAs can be detected once the communication packets are dropped
\cite{ ResMVDC}. \Ared{Cyber-resilient protection schemes for DC microgrids were proposed \cite{ResMVDC,TII_Bipolar_MVDC_LCDRs}. } A Line-model-based scheme to detect FTAs was proposed \cite{Amir1}. Meanwhile, learning-based schemes were proposed in \cite{Ahmadpaper, Kundur_DL,Ahmadpaper_TII}. \Ared{Most of the above methods were designed for detecting FTAs and are, thus, involved  \emph{only after} the LCDR is triggered to trip. Such methods are inherently inapplicable for detecting FMAs.} \Ared{Generally, little attention} was given to securing LCDRs from cyber-induced  FMAs, even though FMAs can lead to serious consequences on the power system by concealing physical faults on the lines protected by the attacked LCDRs \cite{ Ahmadpaper}. 
Prompt and accurate detection of FMAs on LCDRs is, at least, on par with detecting FTAs, which most previous works focused on.
Therefore, \Ablue{there is an urgent need for a robust and accurate solution to detect FMAs targeting LCDRs. Due to the stealthy nature of these attacks, they can lead to catastrophic failures in the power grid. This gap is addressed by the proposed solution as explained next.}

\subsection{ Solution Challenges and Our Contributions}

\subsubsection{Main challenges}

One of the main challenges in proposing a detection mechanism for FMAs is that the proposed solution has to work in real-time, i.e., to detect FMAs within the same time LCDRs would normally take to detect the fault, which is upper-bounded by 1.5 power cycles \cite{GELCDR}. 
The proposed mechanism should also have a very small False Positive Rate (FPR); otherwise, the solution would produce false alarms, resulting in unnecessary service disruption.

\subsubsection{Proposed framework} To tackle these problems, this paper presents a two-stage framework, based on the protected line's equivalent model from the LCDR's viewpoint and artificial neural networks, to detect cyber-induced FMAs on LCDRs, using only measurements available in today's LCDR.
Hence, the proposed framework differs from all schemes that aim to prevent LCDRs from falsely tripping their lines and traditional bad-data detection techniques that detect noise-corrupted measurements. The developed scheme can be implemented in modern microprocessor-based LCDRs as a final line of defense against cyber-induced FMAs. 
A novel aspect of the proposed framework is utilizing a two-stage solution, enabling us to minimize false alarms. The proposed scheme operates in two steps. Firstly, the first module, the physics-based Mismatch Index (MI), is triggered only when an inconsistency arises between the LCDR's local and remote measurements while the LCDR remains untriggered. This discrepancy could indicate a fault masked by a cyberattack or, under certain circumstances, nearby external faults. The second module, Zone-Confirmation Classifier (ZCC), confirms that the detected masked fault lies on the protected line before declaring an FMA. The ZCC uses an Artificial Neural Networks (ANNs) model. It is activated only after the first module is triggered, allowing us to discriminate between the two mentioned scenarios and minimize the false-positive rate.
Another aspect is that both modules' computational complexity is small. Both do not require complex operations or calculations and can run in real time. 
Further, the utilized neural networks in the ZCC rely only on the LCDR's local measurements, making them resilient to adversarial cyberattacks.

\subsubsection{Summary of Contributions}

To capitulate, the main contributions presented in this paper are:

\begin{itemize}
    \item[--] Demonstrating the vulnerability of LCDRs to FMAs.
    \item[--] Developing a two-module framework to detect cyber-induced FMAs on LCDRs.
\end{itemize}

To verify these contributions, the proposed framework's performance is evaluated under:
\begin{itemize}
    \item FMAs masking faults of all types, different locations, and resistances,
    \item  external faults, to ensure the protection's selectivity is maintained under the attacked remote measurement, which was not discussed before in the literature,
    \item  non-fault system dynamics, including generation and capacitor bank switching, to ensure the protective security is restored, and
    \item variations in the power system's operating points and measurement errors such as noise and instrument devices' transients and saturation.
\end{itemize}

Furthermore, 
\begin{itemize}
    \item a comparative analysis is conducted between the proposed framework and existing methods. Notably, only the proposed two-stage framework can accurately detect FMAs without resulting in false alarms, and,
    \item finally, the proposed framework's performance is validated by implementing it in a real-time simulation test bed.
\end{itemize}

In the remainder of this paper, LCDRs' principle of operation and the attack model are described in Section II. 
In Section III, the proposed two-stage framework to detect FMAs on LCDRs is developed. 
The performance of the proposed framework is evaluated in Section IV. 
Additionally, sensitivity analyses are conducted in Section V.
Further, a comparative analysis and real-time simulations are performed in Section VI.
\Ablue{Afterward, a discussion is
performed in Section VII.}
Finally, Section VIII concludes the paper.

\section{Fault-Masking Cyberattacks on LCDRs}

\subsection{ Vulnerability of LCDRs to Cyber-Induced Attacks}

\begin{figure}[b!]
\centering
\includegraphics[width=1\columnwidth]{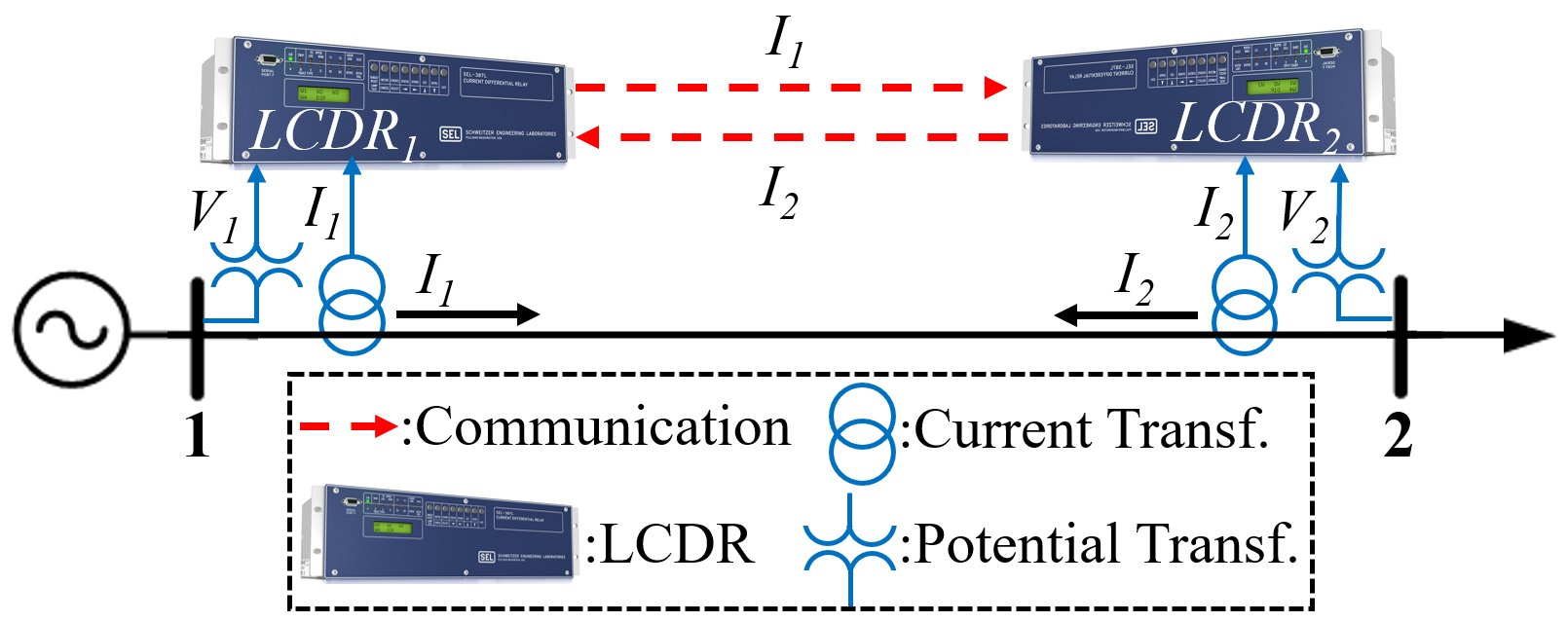}
\caption{
Illustration of line protection by LCDRs.
}
\label{fig:LCDR_idea}
\end{figure}

\begin{figure}[b!]
\centering
\includegraphics[width=0.6\columnwidth]{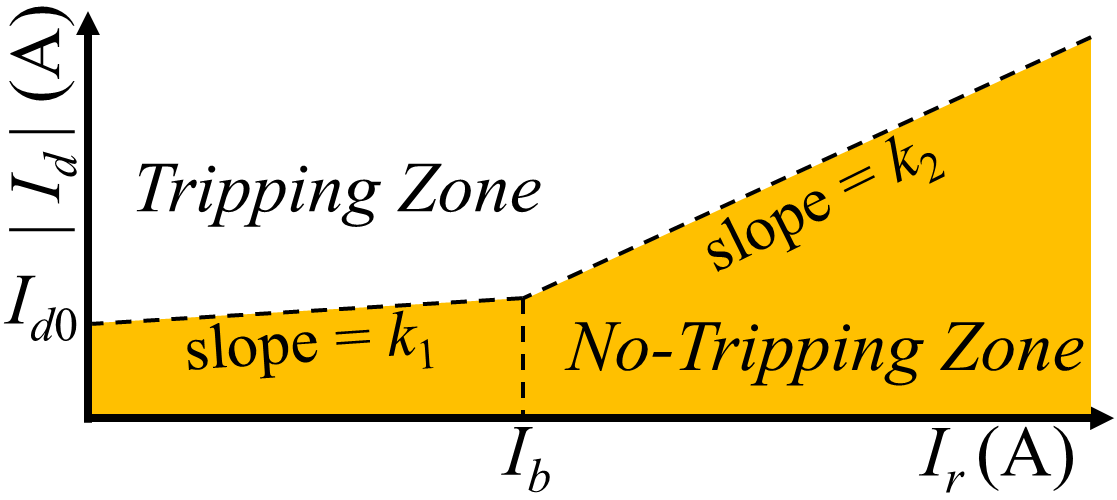}
\caption{LCDR's characteristics.}
\label{fig:LCDR_work_principle}
\end{figure}

LCDRs employ two-way communication links, time-synchronization mechanisms, and various techniques to compensate for any communication-related delays or current transformer saturation to achieve their protective merits. 
LCDRs operate according to Kirchhoff's current law, where vectors of the local and remotely-communicated current measurements, denoted $I_1$ and $I_2$, respectively, are used to monitor the line's condition, as illustrated in Fig. \ref{fig:LCDR_idea}. 
Normally, $I_1$ and $I_2$ are identical, with a slight difference caused by the line charging current. However, under internal faults, a significant mismatch can be noticed between the magnitudes and/or angles of the two measurements for each line phase. In an LCDR, both the differential current ($I_d$) and the operating current ($I_{op}$) are determined as follows:

\begin{equation}
I_d (t) = I_1(t) + I_2(t) 
\end{equation}
\begin{equation}
I_{op}  (t)  =         \\              
\begin{cases}  
   I_{d0} + k_1   I_r (t)   & \text{ } I_r  (t)\leq I_b   \\
    I_{d0} + k_1   I_b + k_2 (I_r  (t) -  I_b)   & \text{ } I_r  (t)\geq I_b 
\end{cases}
\end{equation}

\begin{equation}
 I_r (t)= |I_1(t)| + |I_2(t)|
\end{equation}

\noindent where  $I_{d0}$ is the differential threshold, $I_b$ is the restraining threshold, $I_r$ is the restraining current, $k_1$ and $k_2$ are characteristic slopes, which all shape the LCDR's characteristic lines shown in Fig. \ref{fig:LCDR_work_principle}. The only condition for LCDRs to trip is \cite{GELCDR}:

\begin{equation}
|I_d (t)| \geq I_{op}(t)
\end{equation}

\noindent Before considering cyberattacks, this condition was always met $-$and the LCDR would trip$-$ during internal faults. Now, considering cyberattacks, if $I_2$ is manipulated so that $I_d$ is always less than $I_{op}$, the LCDR's operating point will stay in the blocking region; thus, the LCDR does not trip for internal faults. This cyberattack is the FMA considered in this paper.

\subsection{Attacking LCDRs to Cause Loss of Protection}
Since our work mainly focuses on proposing a defense scheme, it is more practical to assume a powerful adversarial model to ensure that the designed solution protects against attacks from more sophisticated adversaries. 
In particular, we follow the traditional Dolev–Yao model \cite{Dolev}. We consider an adversary with the \emph{Basic attacker capabilities} as described in  \emph{Canetti-Krawczyk} security model \cite{ck_model}.
That is, we assume the attacker ``has full control of the communication link [used by the targeted LCDR]: it can eavesdrop on the transmitted messages [i.e., measurements exchanged between LCDRs], decide what messages will reach their destination [i.e., the targeted LCDR] and when, change these measurements at will or inject its own generated messages'' \cite{ck_model}. We also assume that the communications links are not encrypted (or the adversary can compromise the underlying encryption schemes, e.g., by compromising the endpoints' security and obtaining the encryption keys). Hence, the confidentiality of $I_1$ and $I_2$ is compromised by this adversary.
Naturally, the proposed scheme works for less-powerful adversaries whose manipulations for $I_2$ do not follow the assumed powerful adversarial model described by Equation (5) as explained next. 
Moreover, we assume that the integrity of only $I_2$ can be compromised, unlike local measurements $I_1$ and $V_1$. 
On the one hand, local measurements of an LCDR are measured using the local Current Transformer (CT) and Potential Transformer (PT). Both instrument devices can use direct copper wires or dedicated isolated communication links to send the measurements to the local LCDR \cite{Ahmadpaper}. This allows us to assume, similar to previous works, that local current and voltage measurements ($I_1$ and $V_1$) are secure from manipulation by remote cyberattacks from a remote place, which is the focus of this paper. 
On the other hand, compared to $I_1$, measurement $I_2$ has a wider attack space, making $I_2$ measurement harder to secure.  For instance, attackers can remotely modify $I_2$ after intruding into (i) a vulnerable communication link or (ii) the substation's Local Area Network (LAN) by exploiting the vulnerability of the IEC-61850 standard based on which modern substations' \Ared{communicate.} 
This allows the attacker to exploit this intrusion door (the communication network or the LAN) to eavesdrop, modify, and replace the remote current measurement ($I_2$ as a phasor) \cite{Ahmadpaper, Dolev}. 
Therefore, $I_2$ can no longer be assumed authentic and is assumed vulnerable to cyberattack manipulation.
Finally, for a successful FMA on a certain LCDR, the adversary is assumed to have the necessary knowledge of the working principle of LCDRs, as summarized in the previous sub-section.

\begin{figure}[b!]
\centering
\includegraphics[width=1\columnwidth]{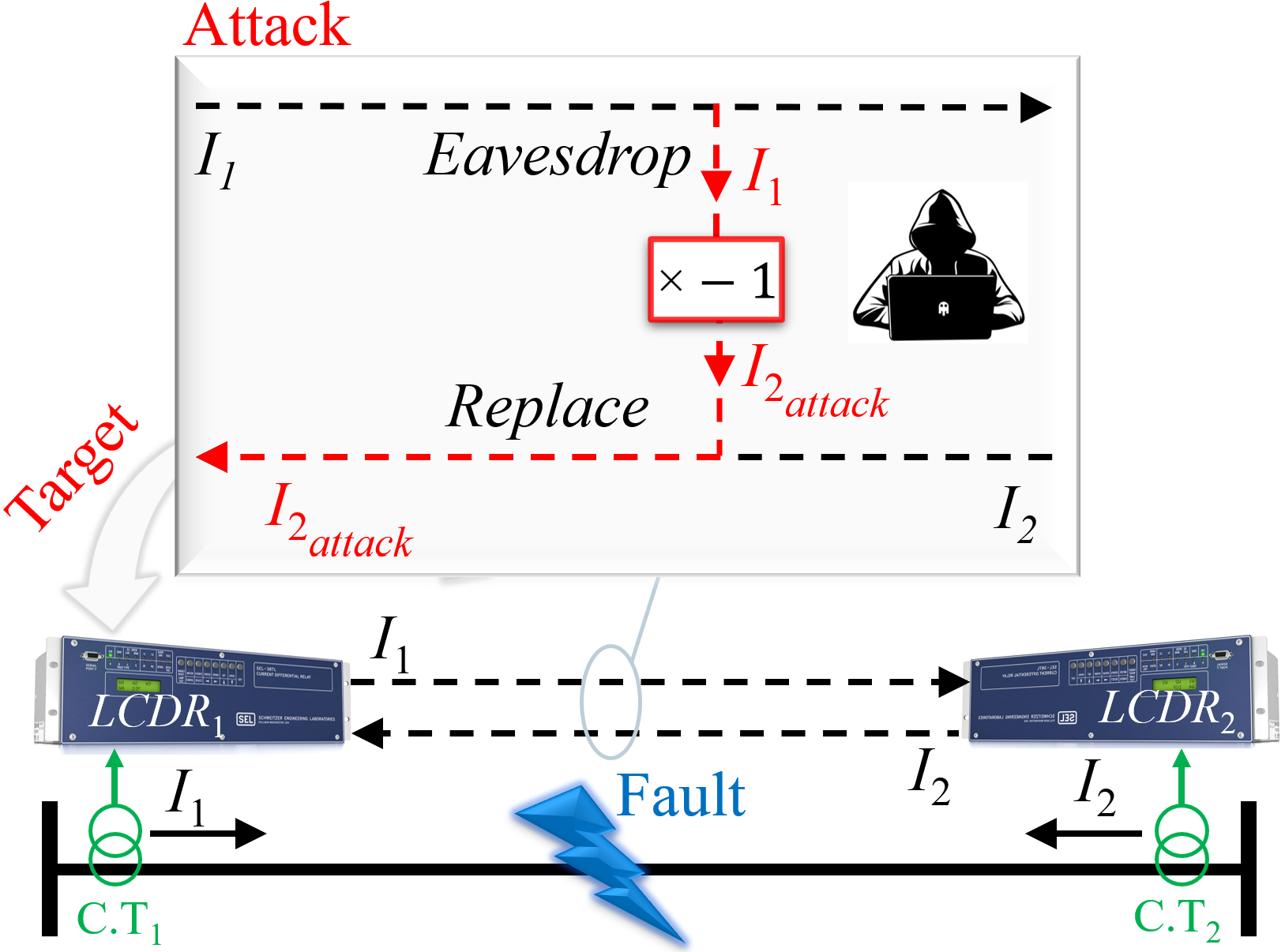}
\caption{Illustration of an FMA on $LCDR_1$.}
\label{fig:attack}
\end{figure}

The FMA is achieved by manipulating the remote measurements sent by the opposite-end LCDR before they reach the attacked LCDR. 
One way to perform an FMA is through a  traditional Man-in-the-Middle (MitM) attack such that the manipulation of the remote measurements starts immediately when a fault is incepted on the line and before the LCDR can sense it. Additionally, the attackers can use malware to manipulate the LCDR's remote measurements. This malware is triggered by any of the signs of fault inception near the LCDR's location, e.g., by the rate of change of the measured current/voltage/impedance near the LCDR, which is a sign of fault inception close to the LCDR's location (but not necessarily on the LCDR's line). The malware can detect this sign in less time than LCDRs take to confirm the existence of an internal fault \cite{Saleh}.  
The attack can also be achieved through a combined cyber-physical attack, where the fault inception time is assumed to be known to the adversary through mutual coordination with the physical-fault initiator \cite{NISTGuidelines, Shahidehpour_Concealing,Shahidehpour_FMA}.
Let us consider attacking $LCDR_1$, as illustrated in Fig. \ref{fig:attack}. 
To perform the attack,  the intruder modifies $I_2$ such that the manipulated remote current measurement received by the hacked LCDR ($I_{2_{man}} $) follows:

 \begin{equation}
I_{2_{man}} (t) = - I_1(t) + C_{a}
\end{equation}

\noindent where $C_{a}$ is a vector that can be arbitrarily set to zero without knowing the system configuration or real-time parameters.   
For a more stealthy attack, the value of $C_{a}$ can be equal to the usual differential current $I_d$ that exists under the normal operation of the line. In this case, $I_d$ can be obtained (i) by eavesdropping the 2-way communication link between the two LCDRs or (ii) from an insider.
By substituting with (5) in Equations (1)$-$(4), it can be concluded that the described attack stealthily masks the true state of the line, and therefore, any fault that occurs on the line will remain undetected. An advantage of the above model is that attackers do not require knowledge of the expected fault current, resistance, or location.
For $LCDR_1$, the remote measurements, $I_2$, sent by $LCDR_2$,
are manipulated, possibly by a MitM network attacker, before they reach $LCDR_1$. For example, the attacker targeting $LCDR_1$ can just drop the $I_2$ measurement coming from $LCDR_2$ and replace it with $-I_1$ in the channel of $LCDR_1$ that is dedicated for $I_2$ measurement.
\Ared{$LCDR_2$, the sender, does not get to see the maliciously modified measurements. 
Nonetheless, if the malicious entity targets $LCDR_2$, i.e., in addition to or instead of $LCDR_1$, the remote measurements of $LCDR_2$, which are now the communicated $I_1$ measurements will need to be manipulated after they are sent by $LCDR_1$ before they reach $LCDR_2$.}

\section{Detecting Fault-Masking Attacks on LCDRs}

Before considering cyberattacks, a line protected by an LCDR would be either healthy or faulty. Considering cyberattacks, confusion arises since unsatisfying equation (4) can either signal a healthy line or one that is under an FMA,  as demonstrated in Section II.
On the one hand, the proposed framework is required to actively monitor the protected line's condition, as long as the LCDR is not triggered, to detect FMAs within the same time the LCDR would have taken to detect the (masked) fault had there been no cyberattack. 
On the other hand, the proposed solution should maintain the LCDR's protective security and zonal selectivity by avoiding maloperation due to external disturbances such as those caused by load/generation switching and faults on adjacent lines.

\begin{figure}[b!]
\centering
\includegraphics[width=1\columnwidth]{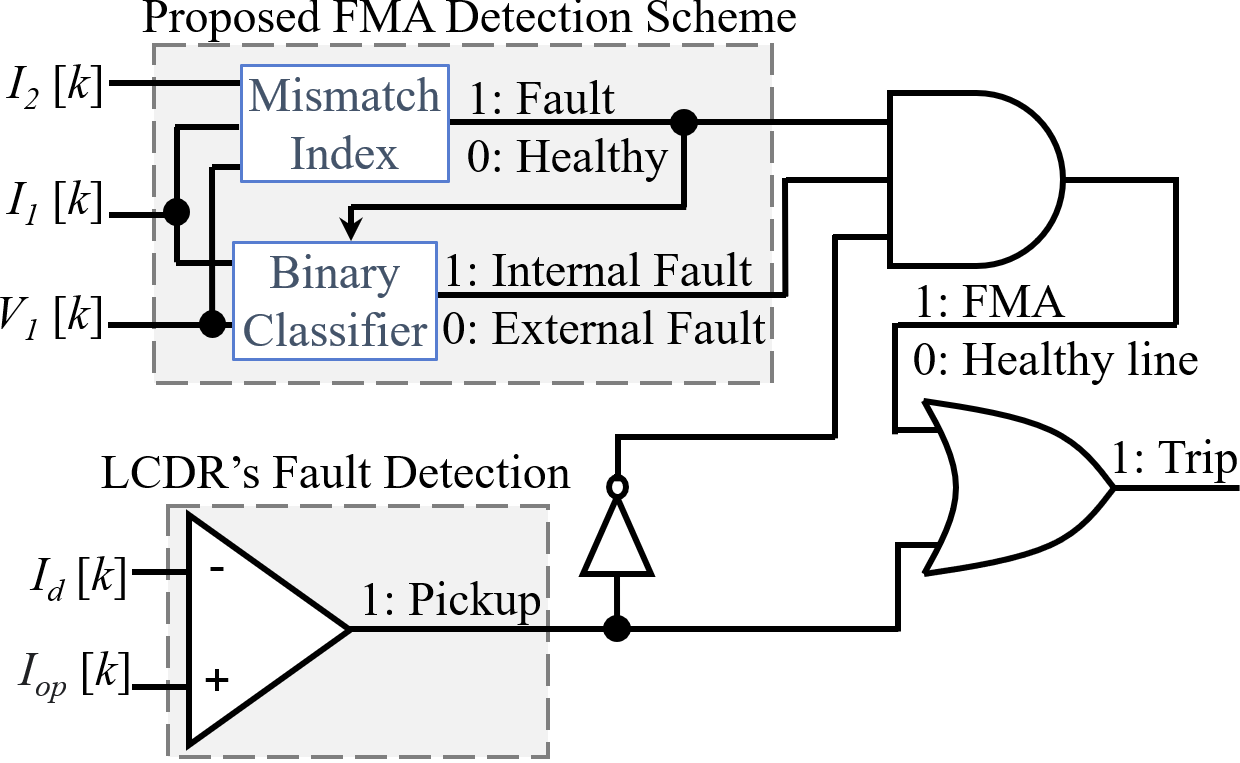}
\caption{ Combined logic of LCDRs with the proposed FMA detection scheme.}
\label{fig:Logic}
\end{figure}

Fig. \ref{fig:Logic} illustrates the LCDR's logic after integrating the proposed FMA-detection module, which can be easily implemented using logic gates. If an FMA is detected, a trip decision can be directly generated (as illustrated). Alternatively, the power system operator can use the FMA alarm to activate other protective relays on the line (backup relays) to double-confirm the tripping decision. The proposed framework is in action only when the LCDR's fault detection module is not triggered. Therefore, the proposed solution can be integrated with any scheme that is designed for detecting FTAs, e.g., the scheme proposed in \cite{Ahmadpaper}, since both schemes work independently and in different instances.
As shown in the figure, the proposed framework only requires the same phasor measurements available for a typical LCDR, which are the locally measured voltage ($V_1$), locally measured current ($I_1$), and the communicated current measurement ($I_2$) \cite{LCDR_faultlocation}. 
When the LCDR is not triggered, the scheme determines if these measurements correspond to the healthy line \Ared{or an FMA.} To perform this task, the proposed framework utilizes two modules: (i) \Ared{the MI, and (ii) ZCC.}
The main role of the MI is to detect mismatches between \emph{the local and remote measurements under FMAs,} which the LCDR confuses with the non-attacked healthy-line measurements. The proposed MI works based on the healthy equivalent model of the protected transmission line and the anticipated values of these measurements under normal operation.
Mismatches detected by the MI are a strong sign of FMAs but can also result from strong faults near the LCDR of concern since these faults affect the utilized measurements, e.g., by depressing the local voltages.
After the MI is triggered, the ZCC confirms that the detected fault falls within the protected line  \emph{using only the local-side current and voltage measurements} since the remote measurements cannot be trusted now. 
If both the MI and ZCC are triggered, an intrusion alarm is automatically raised, and an FMA is declared. Otherwise, the LCDR and the proposed module remain idle. After clearing the detected masked fault, the proposed framework and LCDR are reset.

\subsection{Developing a  Mismatch Index for Detecting FMAs Based on the Transmission Line's Equivalent Model }

\begin{figure}[b!]
\centering
\includegraphics
[width=0.9\columnwidth]
{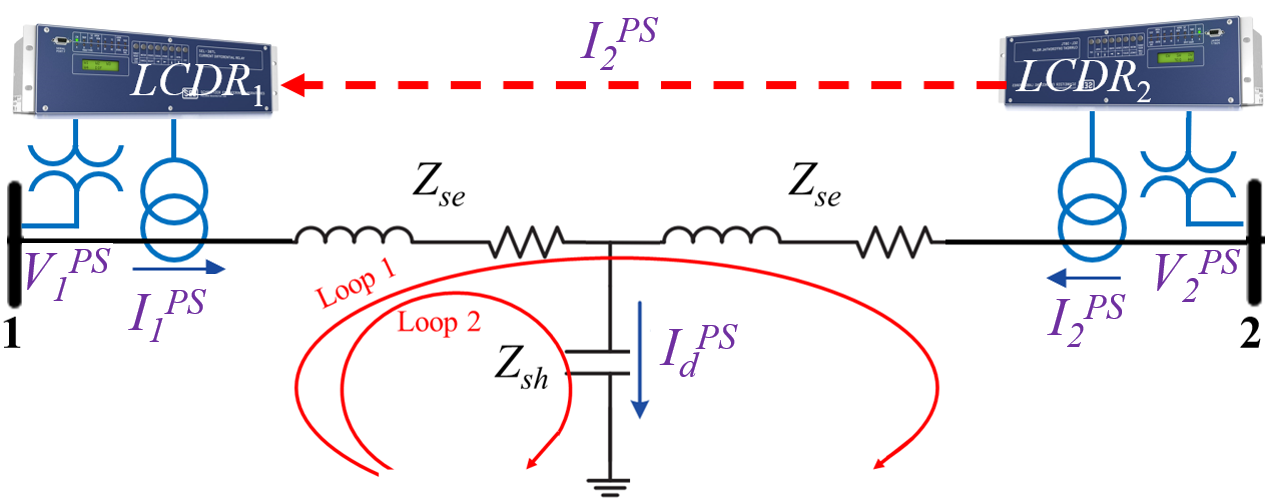}
\newline(a)\\
\vspace{0.2cm}
%
\includegraphics
[width=1\columnwidth]
{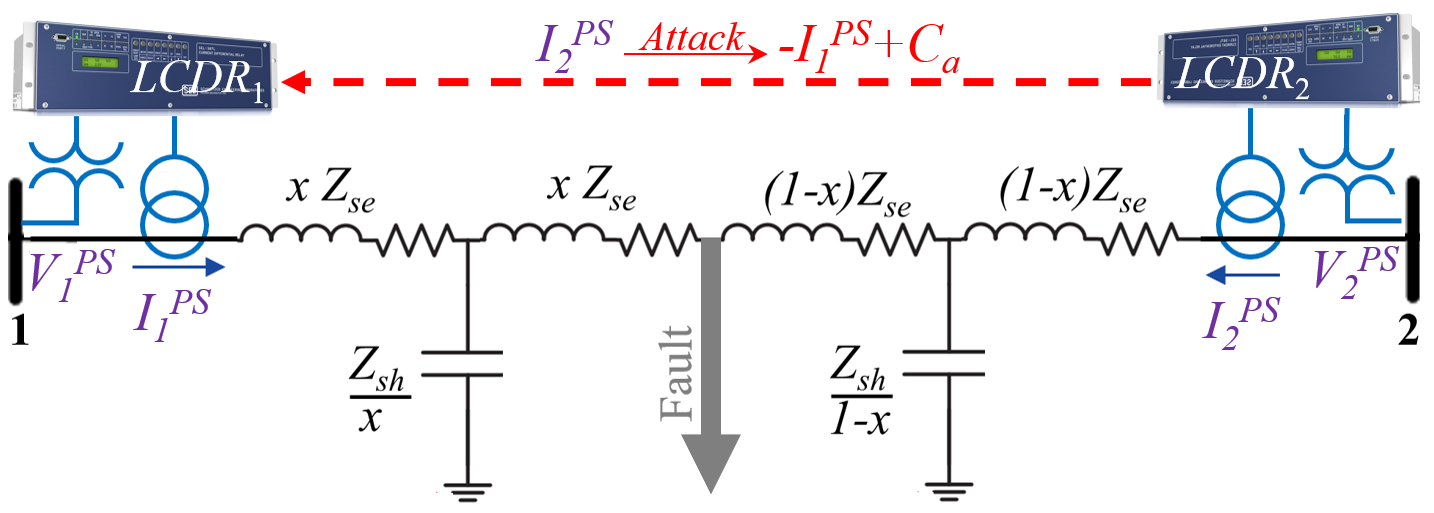}
\newline(b) 
\caption{Line equivalent model.  (a) Healthy,
(b) Under an FMA. 
Concerning $LCDR_1$, measurements $I_1$ and $V_1$ are authentic, but $I_2$ can be manipulated.} 
\label{fig:Equivalent_model}
\end{figure}

Fig. \ref{fig:Equivalent_model} shows the equivalent transmission line models, as viewed by the LCDR, under normal operation and an FMA, illustrated in the Positive Sequence (PS) domain. The fault illustrated in Fig. \ref{fig:Equivalent_model} (b) is located at a distance $x$, expressed as a percentage of the line length and measured from $LCDR_1$. 
Intrinsically, the equivalent circuit of the healthy line has parameters different from those of the faulty line.
In the communication layer, only the remote current measurements can be manipulated by cyber attackers. More details can be found in \cite{faultrate}. The main underlying hypothesis is that cyber-attackers in FMAs cannot manipulate the remote measurements in a way that makes $I_2$,  $I_1$, and  $V_1$ satisfy all equations extracted from the healthy line model due to the aforementioned intrinsic difference between the two physical models of the healthy and the faulty line.
One of the differences between the equivalent circuits of a healthy line (Fig. \ref{fig:Equivalent_model} (a)) and a line under an FMA (Fig. \ref{fig:Equivalent_model} (b)) is that the shunt impedance ($Z_{sh}$) is capacitive in the former case and resistive in the latter \cite{Amir1}.
In the case of an FMA, the measured (true) value of the local voltage ($V_1^m$) is

 \begin{equation}
  V_1^{m} (t) =  I_1 (t) . (2 x  . Z_{se}) + I_d (t) . R_f  
\end{equation}
  
\noindent where $Z_{se}$ is the inductive series impedance of the healthy line and $R_f$ is the fault resistance. 
(In (6), both $I_1$ and $I_d$ are those of the fault, i.e., without manipulation.) The above line model can be used as a basis for detecting FMAs since, for instance, this model can be used to calculate $V_1$ using  $I_1$ and $I_2$ measurements.
Comparing the measured and calculated values of $V_1$ can be used as a security measure to reduce the attacker's degree of freedom since, under FMAs, the difference between these two values will be high. 
To illustrate this point, let us determine how to perform an FMA and bypass this security measure, i.e., how to make the measured $V_1$ equal to $V_1$ calculated from the healthy-line model. 
If LCDRs employ this security measure, the condition of a successful FMA, instead of Eq. (5), becomes:

 \begin{equation}
I_2^{attack} (t) =  ( \frac{   I_1 .2x.Z_{se} + I_d . R_f - I_1 (Z_{se} + Z_{sh}) }{Z_{sh}}   )
\end{equation}

\noindent where $I_2^{attack}$ is the manipulated value of $I_2$. After simplification, (7) leads to:

 \begin{equation}
\frac{I_d^{attack} (t)}  { I_1  (t)} =  (2x-1) Z_{se}
\end{equation}

\begin{equation}
\angle I_d^{attack} (t) - \angle I_1 (t) =
 \begin{cases}
\angle Z_{se}     & x > 0.5 \\
0 & x = 0.5 \\
\angle Z_{se}  + 180^{\circ}  & x<0.5
\end{cases}
\end{equation}

\noindent Equations (7)$-$(9) involve $x$ and/or $R_f$, which are not necessarily known to the attacker. It is worth noting that an FMA can be launched even if the condition in this equation is not satisfied; Eq. (7) is the condition for a successful FMA if LCDRs employ what we propose here, which is to monitor the difference between the measured local voltage and the local voltage calculated from the transmission line's healthy equivalent model. In other words, after LCDRs employ this proposed countermeasure, the only way the cyberattack can bypass this countermeasure without being detected is if ($I_2^{attack}$) follows Eq. (7).
Comparing the measured and calculated local voltages restricts the attacker's degree of freedom and helps detect the FMA. The proposed mismatch index is introduced next. It employs several measures extracted from the healthy-line equivalent model to detect FMAs and differentiate them from the condition of a healthy line. Some of these measures rely on the difference between measured and calculated  $V_1$.

\vspace{3pt} 

\noindent  	• \emph{Proposed Mismatch Index:}
To differentiate between a healthy line and an FMA (illustrated by Fig. \ref{fig:Equivalent_model}; (a) and (b)), the proposed MI (i) employs multiple parameters and equations extracted from the healthy line model, and (ii) uses all the available measurements (i.e., $I_2$, $I_1$ and $V_1$) to check the values of the utilized parameters. 
If the line is healthy, all three measurements for each phase are consistent with the healthy circuitry. Therefore, the mismatch between the calculated and the measured/anticipated values of all the parameters in the MI is negligible. Meanwhile, under FMAs, the local and remote measurements produce a large mismatch in the MI's parameters since only $I_2$ is manipulated while $V_1$ and $I_1$ are of an actual fault. Fig. \ref{fig:Equivalent_model} (a) shows that a healthy line can be represented by two loops which include $Z_{se}$ and $Z_{sh}$.
Three main equations can be obtained from this model. The first equation is equation (1), described in Section II. It is based on Kirchhoff's current law and is the main equation used by LCDRs. 
The two other equations are (6) and (7), which can also be concluded from the model based on Kirchhoff's voltage law. Firstly, for Loop 1 in Fig. \ref{fig:Equivalent_model} (a), the voltage drop ($V_{drop}$) across the line can be determined as

 \begin{equation}
V_{drop} (t) = V_1(t) - V_2(t) =  I_1(t) . Z_{se} + I_2 (t) . Z_{se}
\end{equation}

The absolute value of $V_{drop}$ is continuously monitored and minimized.
However, usually under faults, a sudden increase in the magnitude of $V_{drop}$ can be noticed.   
Secondly, in Loop 2 of Fig. \ref{fig:Equivalent_model} (a) (local-side loop with respect to $LCDR_I$)

 \begin{equation}
V_1 (t) =  I_1(t) . Z_{se} + I_d (t) . Z_{sh}
\end{equation}

Using equation (6), the difference between the calculated and the measured local voltage can be obtained as follows:

 \begin{equation}
\delta_{V_M} (t) =    \frac { |V_{1_{c}}(t) |- |V_{1_{m}} (t) |}   { |V_{1_{m}}(t) |}
\end{equation}

 \begin{equation}
\delta_{V_A} (t) =    \frac { \angle V_{1_{c}} (t) - \angle V_{1_{m}}  (t)}   { \angle V_{1_{m}} (t) }
\end{equation}

\noindent where $V_{1_{c}}$ and $V_{1_{m}}$ are the calculated and measured values of $V_1$, respectively. 
$I_1$ is taken as the reference of all calculations in this paper.
Additionally, $\delta_{V_M}$ is related to the voltage magnitude, while $\delta_{V_A}$ is related to the voltage phase angle.
For a healthy line, both values of $\delta_{V_M} $ and $\delta_{V_A} $ are ideally zero. On the contrary, under FMAs, there will be a big mismatch between the measured and the calculated values of $V_1$ due to the difference in the line model's parameters under faults, which affects both the magnitude and angle of $V_1$. 
From equations (1), (6)$-$(9), $P(t)$ can be defined as

\begin{dmath}
 P (t) =  [ \delta_{V_M} (t), \delta_{V_A} (t),  |V_{drop} (t) |    , 
\frac{|I_d(t) | }{ |I_d^{n}|}     ]
\end{dmath}

\noindent where $I_{d}^n$  is the differential current under the normal operation of the line.
Normally, the mismatch between the measured and the calculated value of $V_1$ is approximately zero, unlike in the case of FMAs. Similarly, the ratio $\frac{|I_d(t) | }{ |I_d^{n}|}  $  is close to 1 when the line is healthy. In some cases, a slight but possible change in the magnitude of $I_d$ might occur, for example, due to line voltage variation under different system loading levels. $|V_{drop}|$ under FMAs (or nearby external faults) is much larger than the case of a healthy system. 
The norm of $P$ is obtained to simplify the selection of the threshold \cite{Euclidean}. 
At any given time, the Euclidean norm of $P$ is

 \begin{equation}
 ||P(t)||_2 =   \sqrt{\sum_{i=1}^{4} p_i^2 (t) }
\end{equation}

\noindent where $p_i$ denotes the individual elements of $P$.
Based on $  ||P(t)||_2$, the Trailing Moving Mean ($M$) can be obtained as

 \begin{equation}
M (t) = \frac{1}{T_1 + 1}  \sum_{i=0}^{T_1}   ||P(t-i)||_2
\end{equation}

\noindent where $T_1$ denotes the number of preceding time points over which the \Ared{$M(t)$} is calculated. For example, when $T_1$ equals $k$, it implies that $M(t)$ is calculated using the measurements received for the current time step and the measurements of the $k-1$ preceding time steps.
 $M$ smooths out minor fluctuations in $P$, such as those caused by measurement noise.
Under normal operation, including healthy dynamics on the system, the rate of change of $M$ is less than that accompanied by faults.
Therefore, a dynamic threshold can be defined beyond which the mismatch in the local and remote measurements is considered associated with a fault in the system. In this paper, the threshold or  healthy upper limit is defined as 

 \begin{equation}
 L_U (t)  =     \frac{1+f}{T_2+1}   \sum_{i=0}^{T_2}   ||P(t-i)||_2 , \mbox{ } T_2 >> T_1
\end{equation}

\noindent where $f$ is a safety factor
that aims to provide a safe vertical distance between ($M$) and the dynamic threshold $L_U$ so that $M$ crosses $L_U$ only in case of a possible masked fault on the system.
$T_2$ is a constant number similar to $T_1$. A discussion on the values of $f$, $T_1$, and $T_2$ comes in Section IV. 
The utilized dynamic threshold $ L_U $ preserves more information on the history of $P$ as it includes more time instances.
Normally, where there is no sudden large mismatch in the parameters of $P$, i.e., no-fault, the two parameters $ L_U $ and $M$ appear as parallel lines, with $L_U$ greater than $M$ by the margin of $f$.
In fault cases, the disruption in $P$ appears more significantly in $M$ than $L_U$ due to the averaging over fewer previous values of $P$ in $M$ than in the threshold $L_U$.
Building on the preceding discussion, the following condition is employed in this paper to trigger the proposed MI for detecting an FMA

 \begin{equation}
 M (t) \ge L_U (t)
\end{equation}

\begin{figure}[t!]
\centering
\includegraphics[width=1\columnwidth]{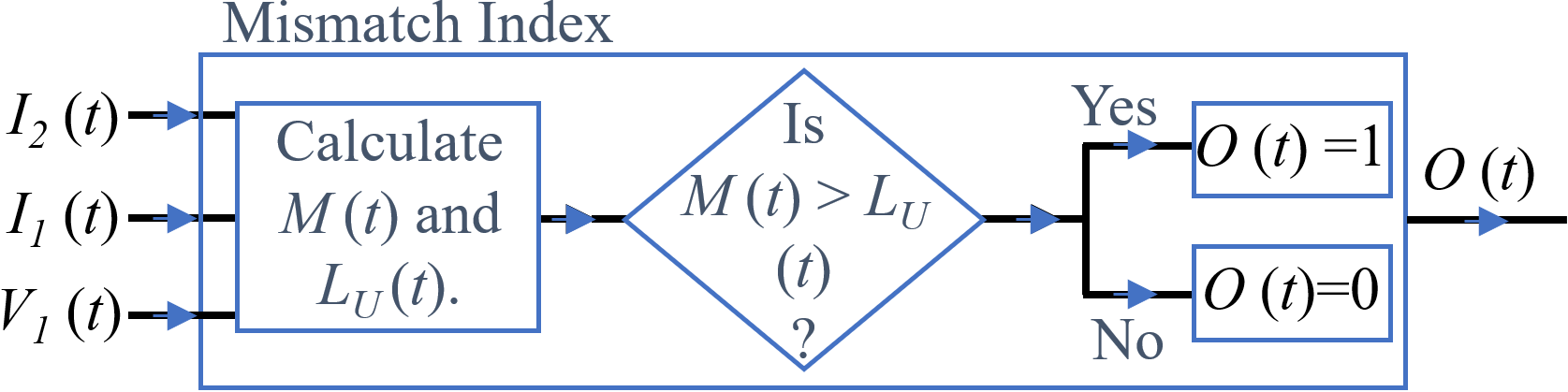}
\caption{Information flow diagram of the  Mismatch Index. 
If $M(t)$ exceeds  $L_U(t)$, a sign of FMAs, $ {O}(t)$ turns 1.}
\label{fig:MismatchIndex}
\end{figure}

\noindent Now, once equation (14) is satisfied, the proposed MI is triggered, and its output, denoted $O$, turns $1$ from 0 and remains $1$ until the FMA is cleared.
Note that the attack detection time will not involve waiting for $T_1$ or $T_2$ steps since the corresponding samples already exist at time t.
Fig. \ref{fig:MismatchIndex} illustrates the information flow for the proposed MI.

\subsection{Learning-Based Classifier for Fault Zone Confirmation}

\begin{figure}[t!]
\centering
\includegraphics[width=0.7\columnwidth]{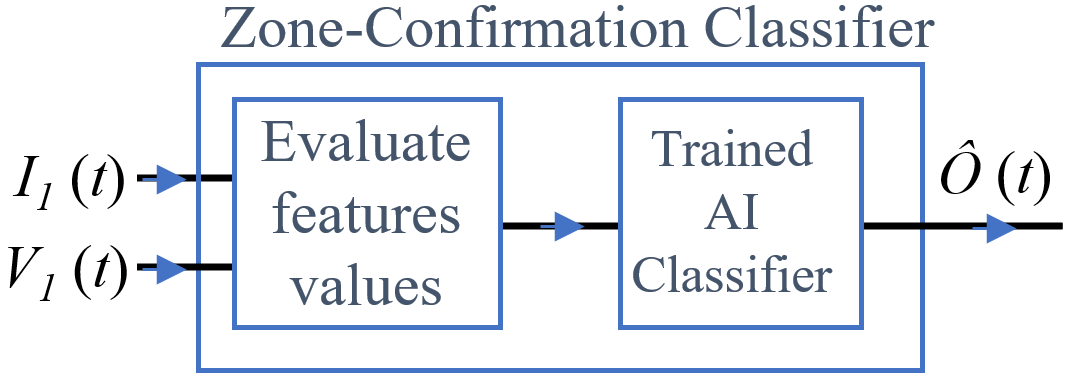}
\caption{Information flow diagram of the Zone-Confirmation Classifier module, activated only after the MI is triggered. 
$\vu*{O}(t)$, the binary output of this module, is 1 under  FMAs and $0$ for external disturbances. 
}
\label{fig:AIClassifier}
\end{figure}

Once the  MI is triggered, confirming that the triggering event lies within the protected line is necessary to ensure the LCDR is still selective, even under cyberattacks.
Therefore, the proposed framework employs
a binary classifier for masked-fault zone confirmation. This module avoids maloperation of the LCDR under severe faults on adjacent lines that may affect $I_1$ and $V_1$ measurements and hence trigger the MI.
Since now that MI is triggered, there is a high possibility that a cyberattack compromised $I_2$ measurement; hence, the ZCC cannot utilize this measurement and has to rely on local-side measurements ($I_1$ and $V_1$).
Additionally, the ZCC must act in no more than a few milliseconds so that the overall operating time of the proposed framework remains within the same time frame usually taken by LCDRs to detect internal faults \cite{GELCDR}.
Therefore, techniques like \cite{LCDR_faultlocation, Takagi}, which take more time than the LCDRs' fault detection time, rely on remote currents, or are not suitable for transmission systems, cannot be used as a part of the proposed solution. 
Learning-based techniques have been previously used for similar purposes, such as fault detection and location \cite{ ANN2}. Inspired by these works, the ZCC is developed as a learning-based module in this paper.
For this purpose, any accurate and fast learning-based classifier can be utilized.
The chosen classifier is first trained offline on two distinct classes of faults: internal and external faults on adjacent lines on both sides. For practicality, the training datasets must cover all the possible fault types at different locations on each line and with various possible fault resistances. 
Next, the trained classifier model can be deployed to differentiate between internal masked and non-internal faults online, i.e., after the MI is triggered. 
The diagram in Fig. \ref{fig:AIClassifier} illustrates the information flow in the proposed classifier. Similar to the MI, the output of the Classifier, denoted $\vu*{O}$, is either 1 (for masked internal faults) or 0 (for external faults). In this paper, ANNs are selected for the classifier due to their accuracy and speed merits. ANNs have been widely used for different classification applications. Offline, ANNs are trained on the datasets of both classes, i.e., in-zone masked faults and out-of-zone disturbances. Online, a trained ANN predicts to which class the newly received data belongs, as explained in \cite{ ANN2}.

\vspace{3pt}

\noindent  	• \emph{Classification Features:} \Ared{After the Mismatch Index is triggered, the classifier extracts the features summarized in Table \ref{table:classification_features} from the non-manipulatable local-side voltage and current measurements for three phases of the line protected by the LCDR.}
In the utilized features,  the superscript ${^P}$ denotes per-phase values, which are fed for the three phases (A, B, and C), and the superscript $^{Seq}$ denotes sequence-domain features, which are fed for the positive, negative, and zero sequences.
$|.|$, $\angle .$,  $Re (.)$ and $Im (.)$ denote the magnitude, phase angle, real part, and imaginary part  
for the corresponding measurement, respectively.
The utilized features include each phase's current and voltage  (magnitude and angle), impedance-related features, and features in the symmetrical components domain. Additionally, the measurements' pre-triggering and post-triggering values are used to represent the fault-imposed change in the measured local currents and voltages.  Note that for large power systems, the system's loading level has a considerably small impact on the local voltage and current measurements compared to the impact caused by the fault itself \cite{faultrate}. Most of the utilized features are available and calculated by modern multi-function relays, such as \cite{GELCDR}, so almost no extra processing power is required.

\begin{table}[t]
\centering
\begingroup
\caption{ Utilized Classification Features}
\begin{tabular}{c | c }\hline
\centering
\makebox{Feature No.}
&\makebox{Feature }
\\   \hline 
\rule{0pt}{2.5ex}     1$-$8 &
$ |V_1^P|,  |I_1^P|,   |V_1^{Seq}|,  |I_1^{Seq}|,$   
$\angle V_1^P,  \angle I_1^P, 
\angle V_1^{Seq},  \angle I_1^{Seq},$
\\
%
\rule{0pt}{4ex} 9, 10 &   
$\angle V_1^{P} - \angle I_1^{P}$, 
$\angle V_1^{Seq} - \angle I_1^{Seq}$, 
 \\ 
%
\rule{0pt}{4ex} 11$-$14 &
$ | \frac{V_1^{P}}{I_1^{P}}|,
\angle  \frac{V_1^{P}}{I_1^{P}},  $
$ | \frac{V_1^{Seq}}{I_1^{Seq}}|, 
\angle  \frac{V_1^{Seq}}{I_1^{Seq}},  $
\\ 
%
\rule{0pt}{4ex} 15$-$18 &  
$Re (\frac{V_1^P}{I_1^P}), 
Im (\frac{V_1^P}{I_1^P}),$ 
$Re (\frac{V_1^{Seq}}{I_1^{Seq}}), 
 Im (\frac{V_1^{Seq}}{I_1^{Seq}}) $
\\ \hline 
\end{tabular}
\label{table:classification_features}
\endgroup
\end{table}

\begin{figure}[t!]
\centering
\includegraphics[width=1\columnwidth]{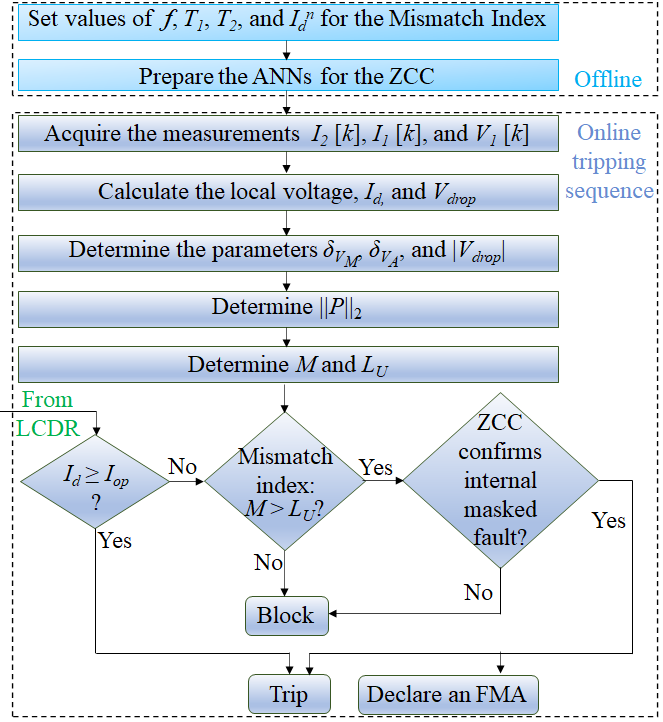}
\caption{Flowchart of the proposed framework.}
\label{fig:flowchart}
\end{figure}

A flowchart for the proposed solution is depicted in Fig. \ref{fig:flowchart}.

\section{Performance Evaluation}

\subsection{ Comprehensive Cases Studies}

\begin{figure}[t!]
\centering
\includegraphics[width=1.0\columnwidth]{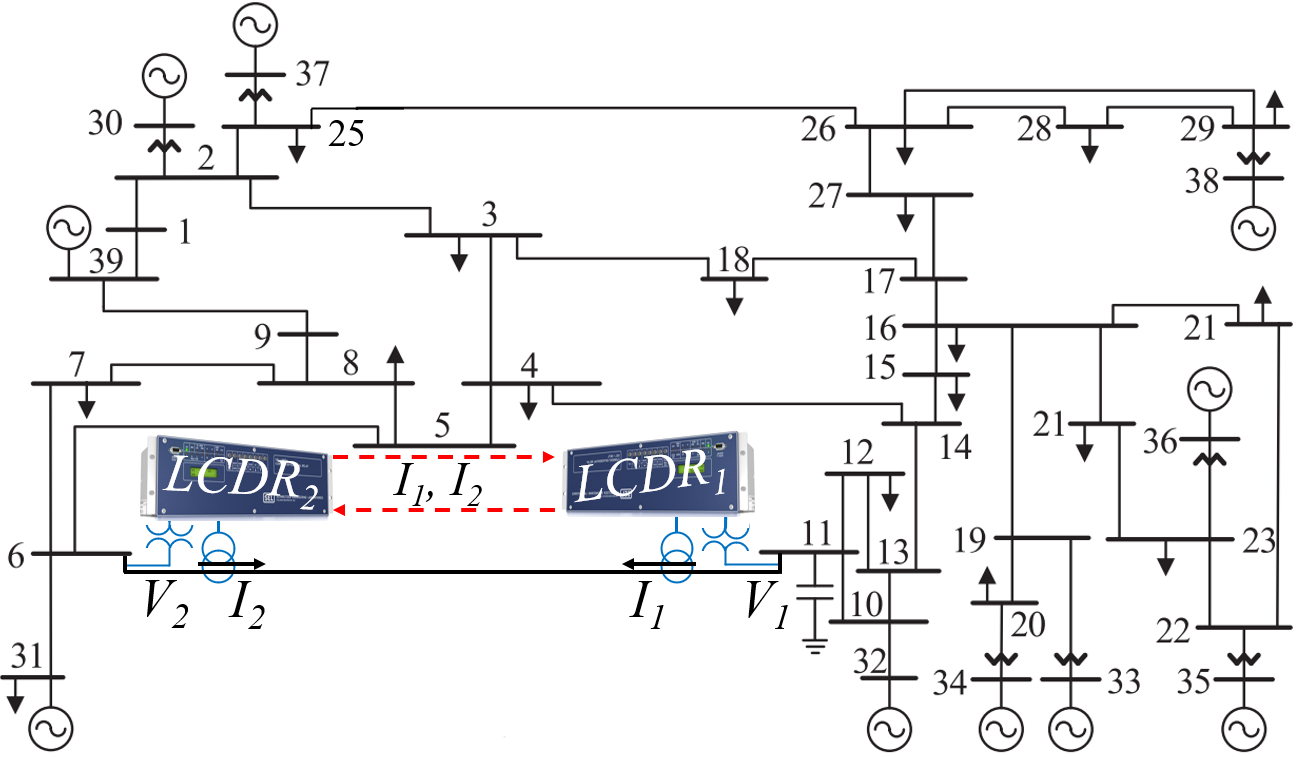}
\caption{The IEEE 39-bus system used in this study.}
\label{fig:39bus}
\end{figure}

The IEEE 39-bus benchmark system, also known as the 10-machine New England power system, is used in this paper to demonstrate the proposed framework, as illustrated in Fig. \ref{fig:39bus}. The system, whose details can be found in \cite{39bus}, is simulated in PSCAD/EMTDC environment.   In this system, the 100-km critical line 11-6 is protected by a pair of LCDRs denoted as $LCDR_1$ and $LCDR_2$, respectively. In this paper,  we focus on $LCDR_1$ to show the performance of the proposed framework.  As shown in Fig. \ref{fig:39bus}, $LCDR_1$, located near bus 11, is locally fed by the measurements $V_{1}$ and $I_{1}$, and receives the current measurement $I_2$ 
communicated from $LCDR_2$ that is near bus 6. Both LCDRs are configured so that $I_{d0}$ and $I_b$ equal to 0.05 kA and 0.585 kA, while $k_1$ and $k_2$ equal to 0.2 and 0.4, respectively, following \cite{Ahmadpaper}. Normally, $0.273 \angle 97.8 ^{\circ} $ kA and $0.29\angle $-$73.56^{\circ} $ kA flow into the line from bus 6 and bus 11, at $ 141.963  \angle $-$100.36 $  kV and $ 138.103 \angle $-$94.72 $  kV, respectively,  so $I_d^{n}$ =   $0.046 \angle  $-$10.02 ^{\circ} $  kA.

During FMAs, attackers intrude into the 2-way communication line between $LCDR_1$ and $LCDR_2$ and manipulate the remote measurements so that each LCDR continuously receives approximately the negative value of its locally measured current in the form of the remotely-communicated current measurement before, during, and after the masked fault's inception, as explained in Section II. Using this test system, two distinct datasets are generated. The first is a comprehensive FMAs dataset, and the second dataset is for external disturbances. To ensure most fault models are represented in the FMAs dataset, faults of all types, different locations, and various resistances are simulated on line 11-6. Next, using equation (5), the FMA cases are generated from the simulated fault cases. 
To vary faults, a wide spectrum of faults is simulated, including A-G, B-G, C-G, A-B, B-C, C-A, A-B-G,  B-C-G, C-A-G, A-B-C, and A-B-C-G faults are simulated at 10\%, 20\%, 30\%,  40\%, 50\%, 60\%, 70\%, 80\%, 90\% of the line, with $R_f$ values of 0.001, 1, 2, 3, 4, 5, 6, 7, 8, 9, 10, 15, 20, 25, 30, 35, 40, 50, 60, 70, 80, 90, 100, 150, 200, 250 and 300 $\Omega$, following \cite{ Ahmadpaper, ANN2}. A total of 5,346 FMAs are generated. To generate the second dataset, 5000 faults of different types, locations, and resistances are simulated on the 5 lines adjacent to line 11-6. Each point in the datasets is sampled at 1K sample/sec, and the utilized features are then determined. To model measurement error cases, measurements are combined with additive, white, and Gaussian noise with 35 dB or higher Signal-to-Noise Ratio (SNR) \cite{Ahmadpaper}.

\subsubsection{ Preparing the Mismatch Index by selecting $f$, $T_1$ and $T_2$} 

\begin{figure}[t!]
\centering
\includegraphics[width=1\columnwidth]
{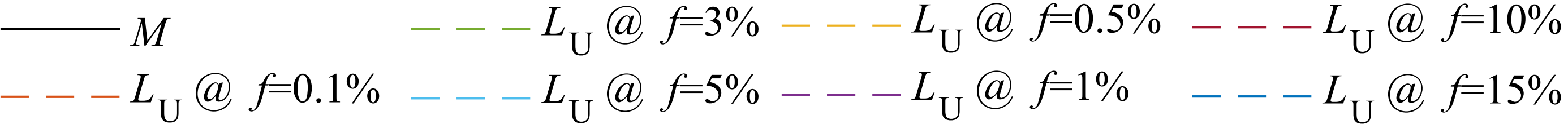}
\includegraphics[width=0.6\columnwidth]
{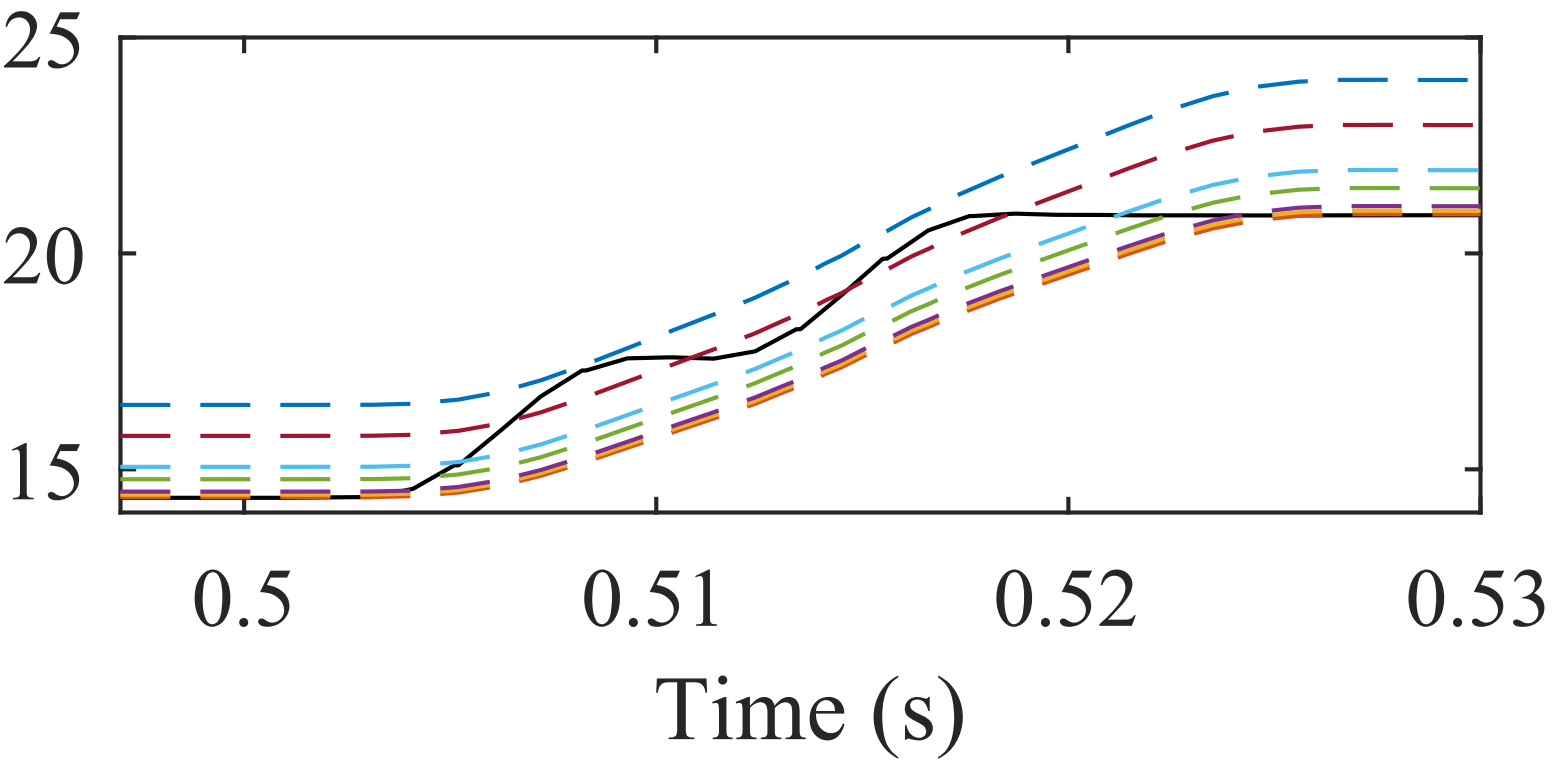}
\\ (a)\\
\vspace{9pt}
\includegraphics[width=0.6\columnwidth]
{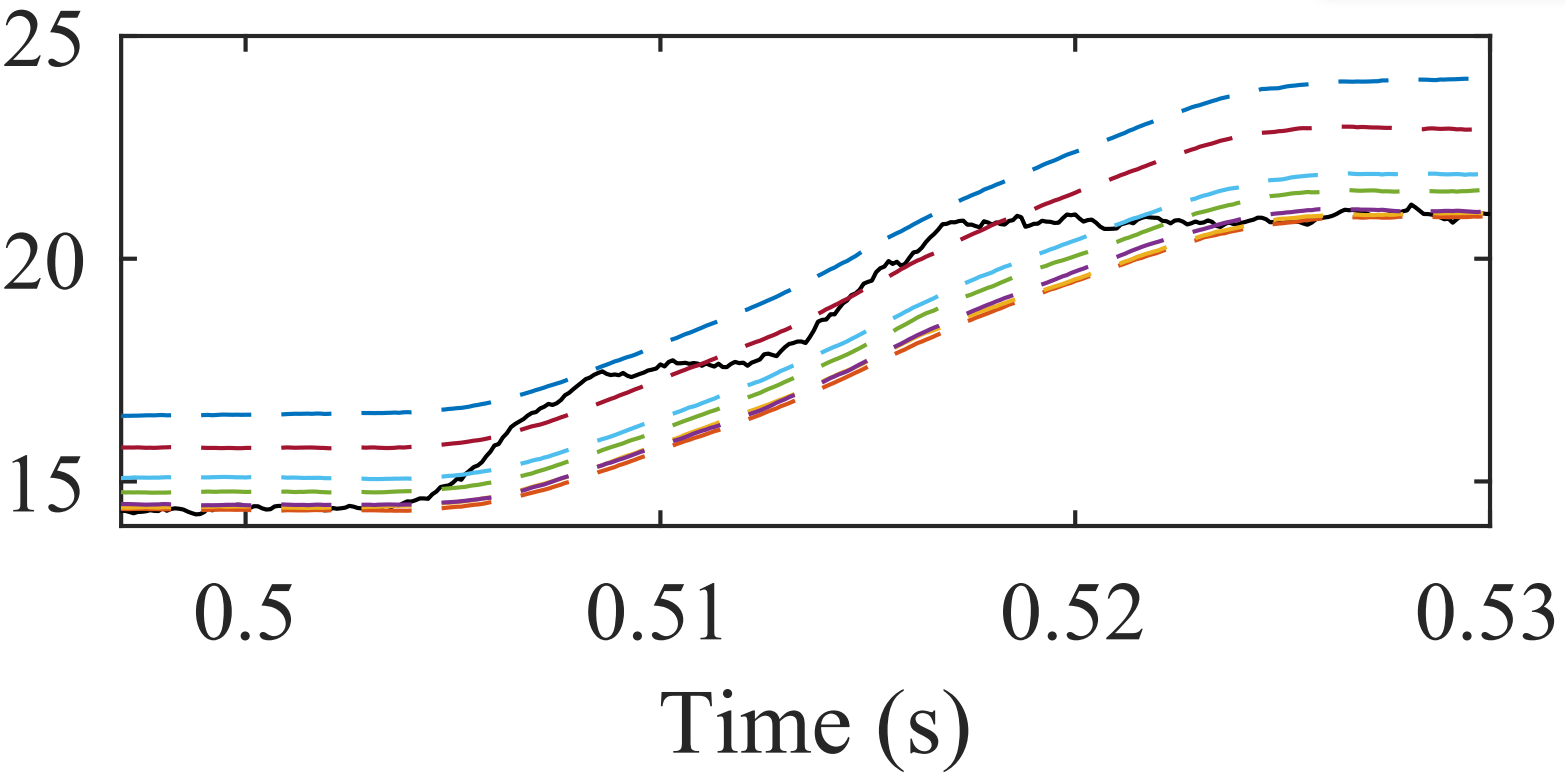}
\\(b)
\caption{Mismatch Index under different $f$s, (a) without noise, (b) with noise.}
\label{fig:f_sensetivity}
\end{figure}

Selecting a value for $f$ is a trade-off between maximizing the rate of true positives (cyberattacks to be detected) and minimizing the rate of false positive cases (false alarms). On the one hand, assigning a large value for $f$ may result in many cases of missed cyberattacks. On the other hand, choosing a very small value of $f$, even though will guarantee that most cyberattacks will be detected $-$even those masking high-impedance faults$-$, may result in many cases of false alarms, e.g., false alarms due to measurement noise under normal operation of the power system.
To illustrate this point, let us consider a case of a cyberattack masking an A-G fault whose resistance is 150 $\Omega$ and located in the middle of the line. 
Different values of $f$ are simulated, resulting in different upper limits ($L_U$s), as depicted in Fig. \ref{fig:f_sensetivity}.
 From the figure, it can be noticed that the mismatch index is (i) robust to variations in $f$ between 3\% and 10\%, (ii) might generate a false alarm under normal operation when $f$ is equal to and below 1\%, and (iii) fails to detect the cyberattack when $f$ is as large as 15\%. 
Moreover, it can also be noticed that increasing the value of $f$ increases the detection time.
In this study, $f$ is selected as 5\%, based on the above discussion.
Similarly, equations (16) and (17) employ moving averages, which involve $T_1$ and $T_2$, to minimize false alarms. In other words, the moving average eliminates minor fluctuations in both $M$ and $L_U$, e.g., fluctuations resulting from measurements' noise, allowing both $M$ and $L_U$ to reflect mismatches resulting from only fault-like system dynamics.  
$T_1$ is selected as 9, corresponding to a 10-point moving average, i.e., approximately one-sixth of a cycle for the LCDR used in this study. A smaller value will result in more fluctuations in $M$, which, in turn, may increase the rate of false alarms. On the other hand, larger values of $T_1$ will increase the time to detect the cyberattack and increase the required computing memory.  
Furthermore, $T_2$ is selected much larger than $T_1$ so that $M$ only crosses $L_U$ under masked faults and not under non-faulty power-system dynamics, e.g., switching a nearby capacitor bank, as shown in the sensitivity analysis section. $T_2$ is selected as 99, which means a 100-point moving average is used to calculate $L_U $, corresponding to averaging over 100 ms.

\subsubsection{ ANN Model Preparation} 

The generated data is labeled, shuffled, and split into 70\% (for training and validation) and 30\% (for testing). Using MATLAB, the hyper-parameters of the ANN (parameters whose values are defined during the training phase) are obtained with the aid of the Bayesian Optimizer (BO) from \cite{Bayes}. By treating the ANN model's hyper-parameters
as optimization variables, the BO aims to maximize the model's accuracy over a series of epochs.  In each epoch, K-fold cross-validation is performed \cite{Ahmadpaper}. In detail, the training and validation dataset is split into K equal folds, where one fold is left out for validation, and the remaining folds are used to train the model. After training the model, it is validated (tested) on the left-out fold, and this process is repeated k times, where, each time, a different fold is used as the validation fold. After k-time training-validation sessions, the model with the highest validation accuracy is selected, and its hyperparameters are considered optimal in this epoch. In this paper, 10-fold cross-validation is performed. One thousand epochs are performed, with iterations limited to 1000 in each epoch, following \cite{Ahmadpaper}. Finally, before the training commences, the misclassification cost is adjusted so that the penalty for false positives, i.e., non-cyberattacks misclassified as cyberattacks, is ten times the penalty for $FN$.  This additional step ensures that the proposed framework does not generate false alarms for non-cyberattack cases. The resultant ANN model after training has 2 ReLU-activated dense layers with 310 and 90 nodes, respectively, a regularization strength of $9.8838\times10^{-7}$, and a Softmax-activated output layer \cite{NNbook}.

\subsection{Performance Metrics and Evaluation  Criteria}

In the testing phase and to evaluate the performance of the proposed framework, 
the following metrics are used \cite{Ahmadpaper}:

 \begin{equation}
accuracy =  \frac{ TP + TN  } { TP + TN + FP + FN  }
\end{equation}

 \begin{equation*}
precision=  \frac{ TP   } { TP +  FP  } \textcolor{white}{..} (20) \textcolor{white}{......}
recall=  \frac{ TP    } { TP  + FN  } \textcolor{white}{..} (21)
\end{equation*}

 \begin{equation*}
FN rate =  \frac{ FN   } { FN +  TP  } \textcolor{white}{..} (22) \textcolor{white}{......}
FP rate =  \frac{ FP   } { FP +  TN  } \textcolor{white}{..} (23)
\end{equation*}

\noindent where $TP$, $TN$, $FP$, and $FN$ denote the True Positive, True Negative, False Positive, and False Negative cases, respectively. In this paper, $TP$ cases are FMAs detected by the proposed framework within 1.5 to 2 power cycles, the time window modern LCDRs take to detect faults \cite{GELCDR}. Correctly-identified non-FMA cases, e.g., healthy fluctuations and external disturbances, are the $TN$ cases. 
Maximizing $TP$ and minimizing $FP$ are both required to regain the LCDR's protective sensitivity, selectivity, and security.

\subsection{Results Discussion}

\subsubsection{Performance Results when Using the Mismatch Index Only}

\begin{figure}[t!]
\centering
\includegraphics[width=0.45\columnwidth]{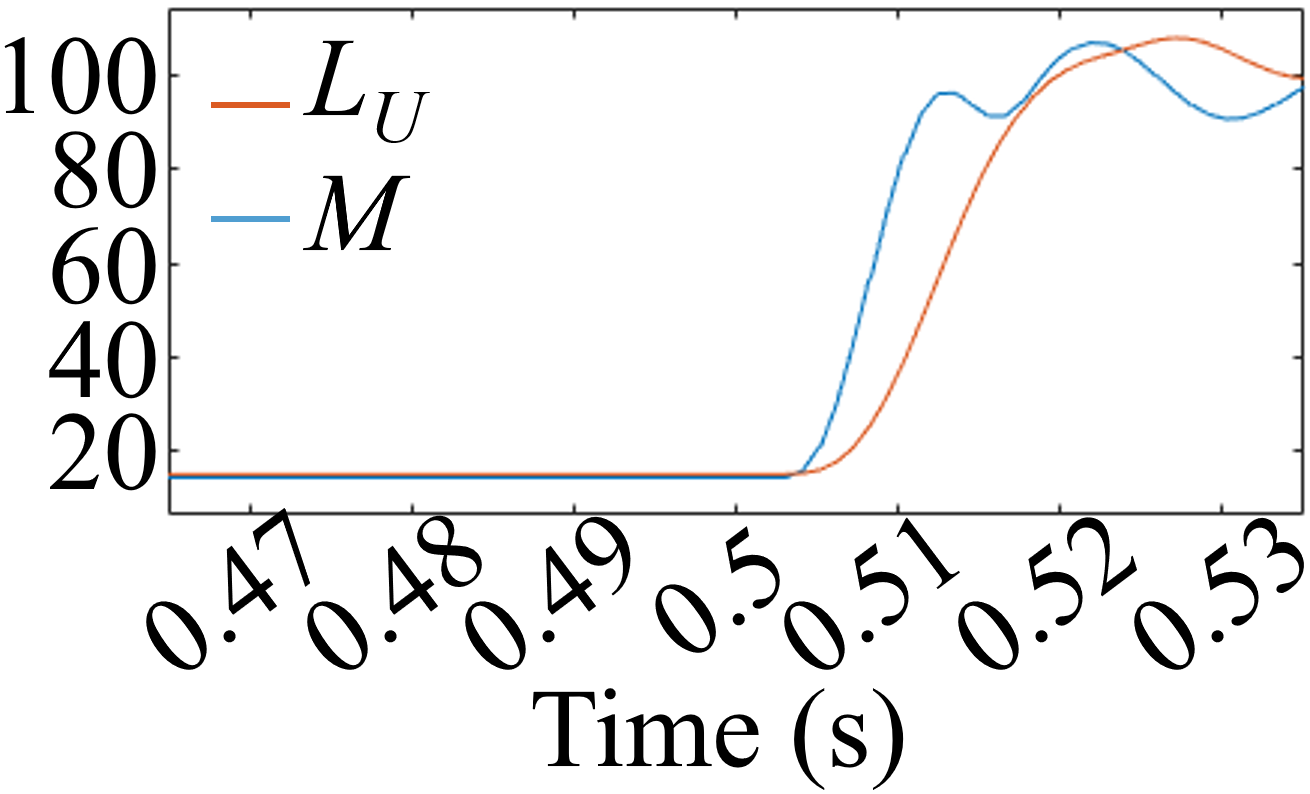}
\hspace{3pt}
\includegraphics[width=0.45\columnwidth]{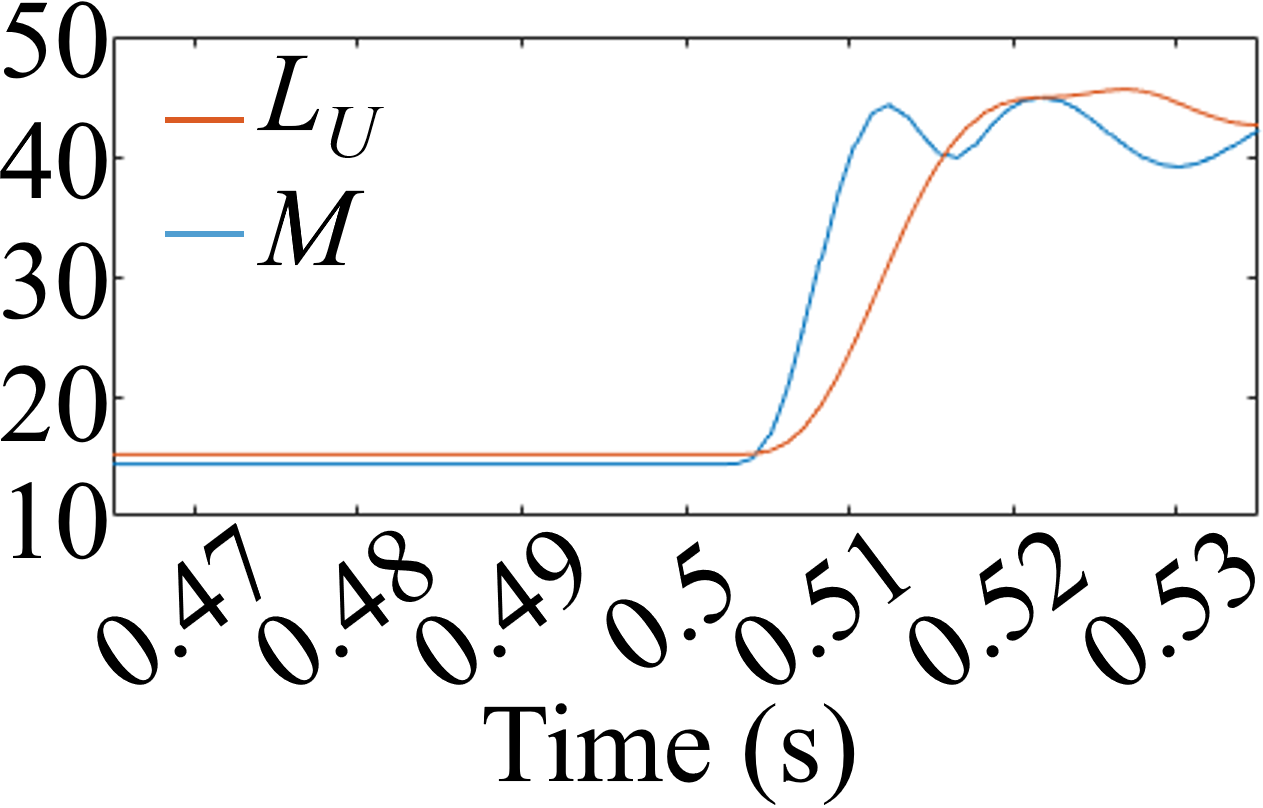}
\\
\hspace{4pt} (a) \hspace{102pt} (b) 
\\ \vspace{5pt}
\includegraphics[width=0.45\columnwidth]{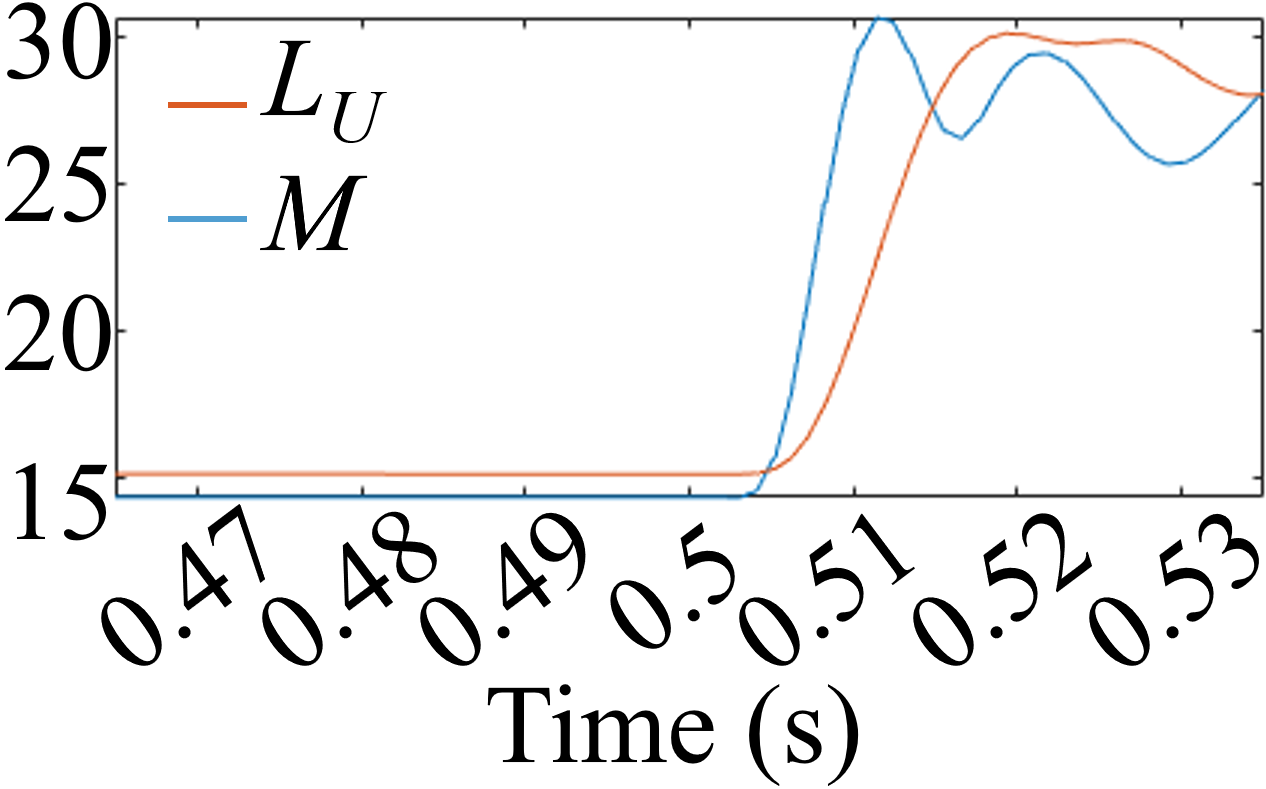}
\hspace{3pt}
\includegraphics[width=0.45\columnwidth]{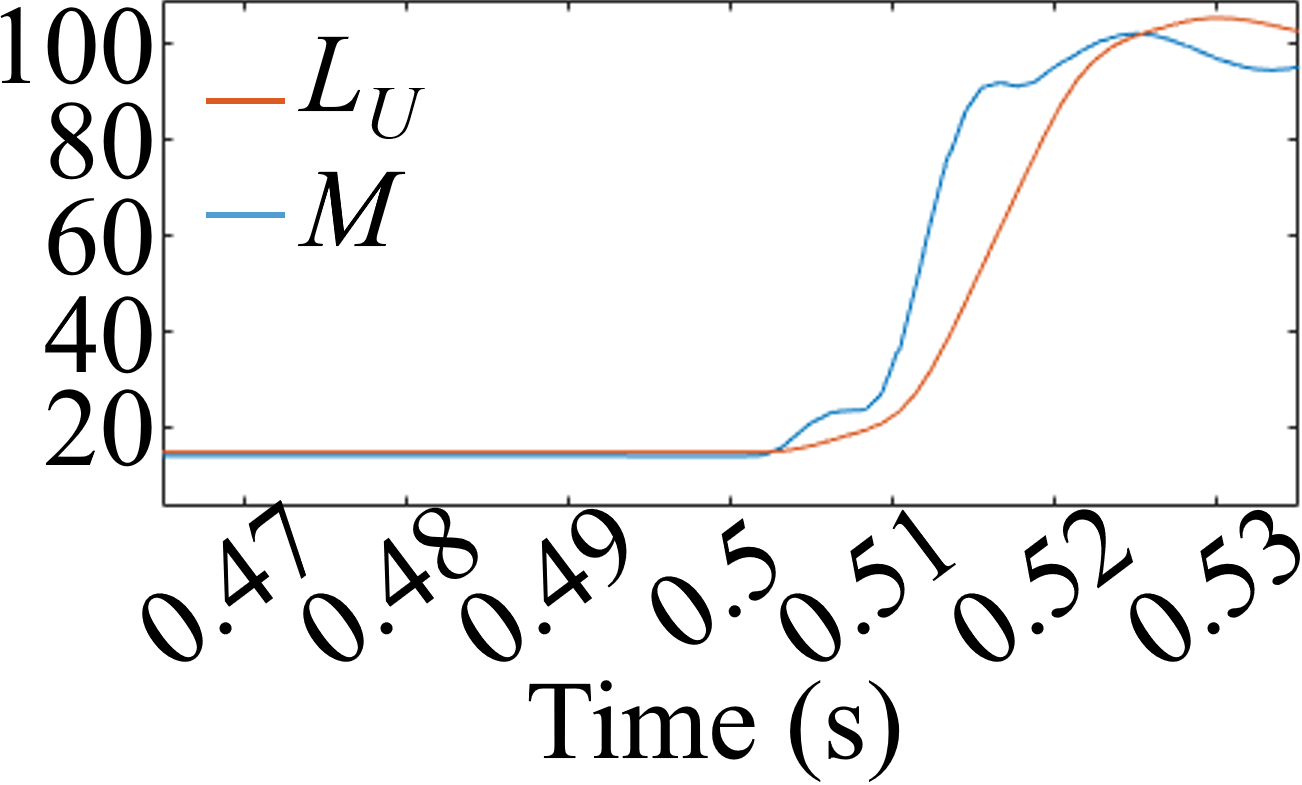}
\\
\hspace{4pt} (c) \hspace{102pt} (d) 
\\ \vspace{5pt}
\includegraphics[width=0.45\columnwidth]{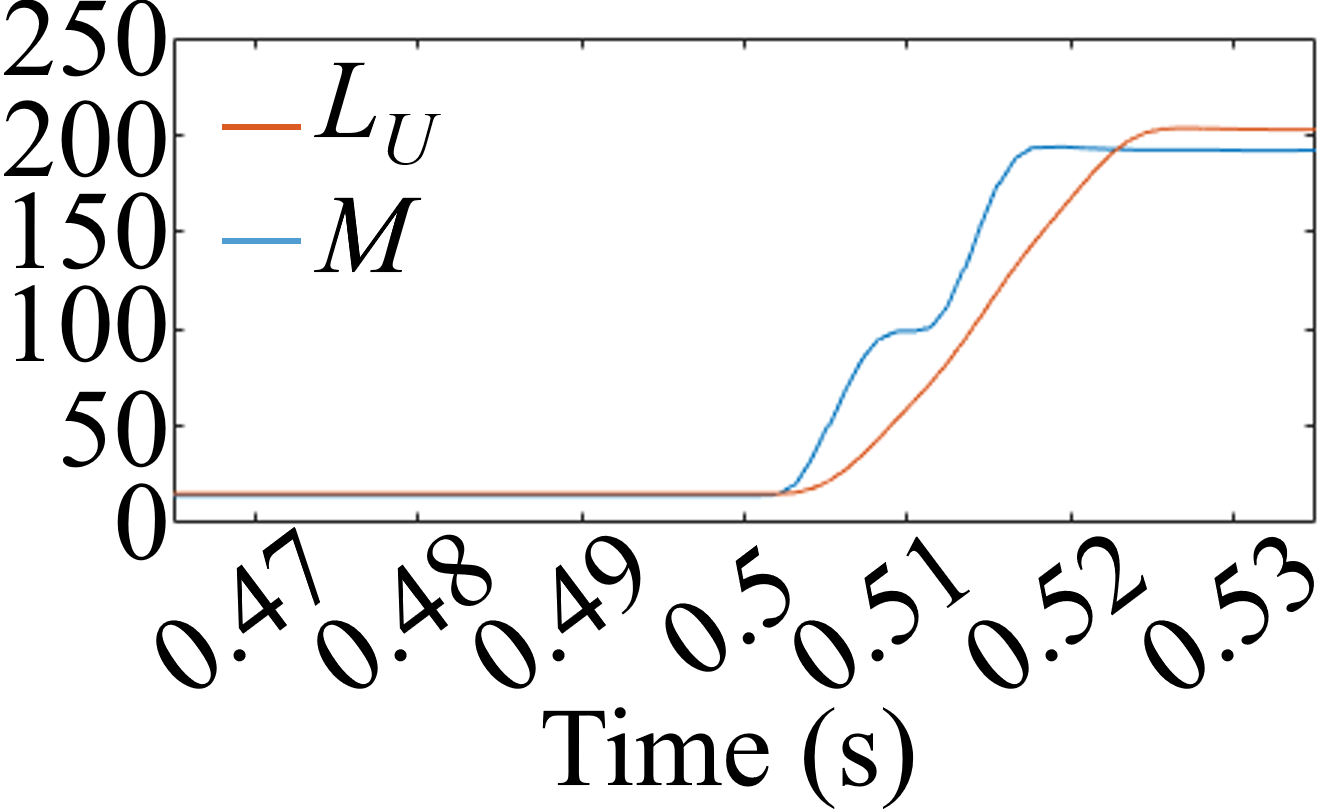}
\hspace{3pt}
\includegraphics[width=0.45\columnwidth]{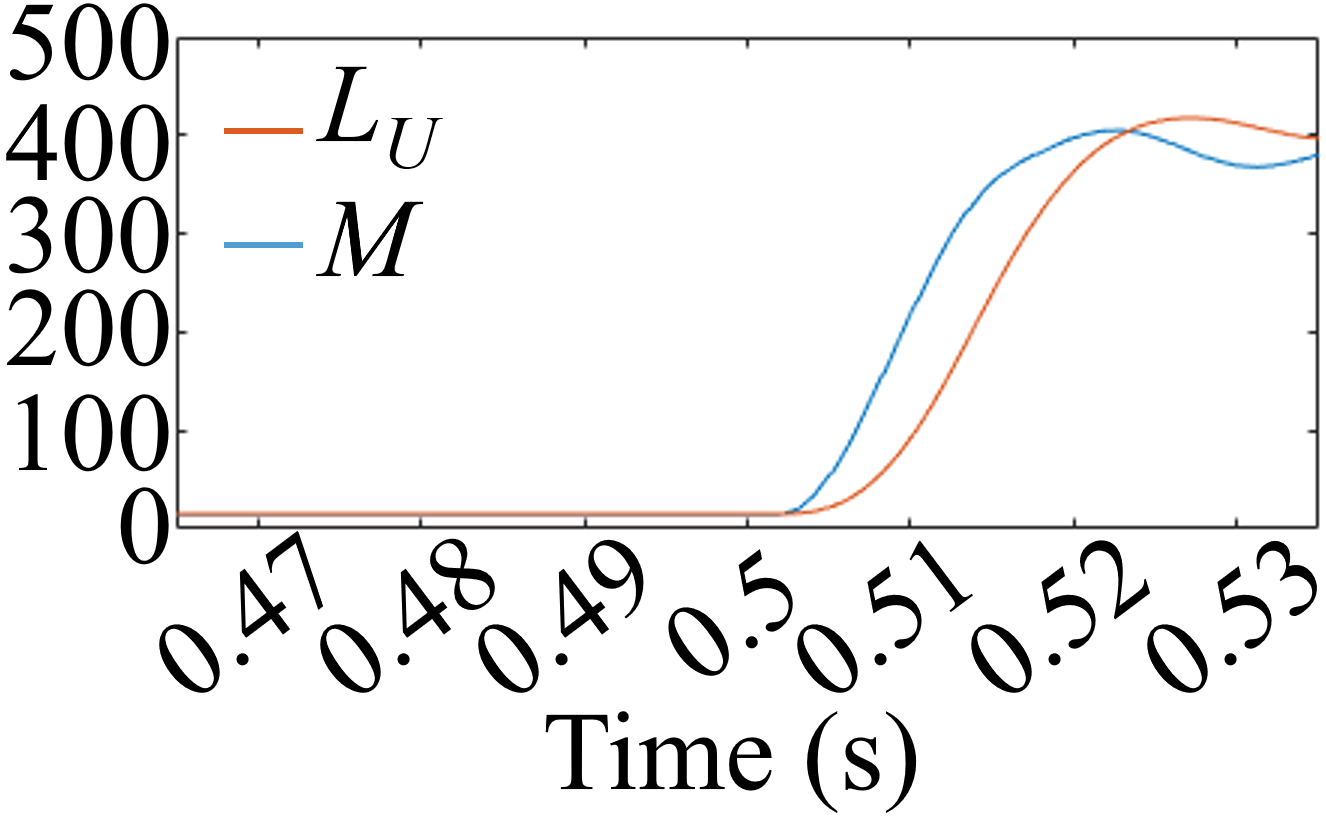}
\\
\hspace{4pt} (e) \hspace{102pt} (f)
\\ \vspace{5pt}
\includegraphics[width=0.45\columnwidth]{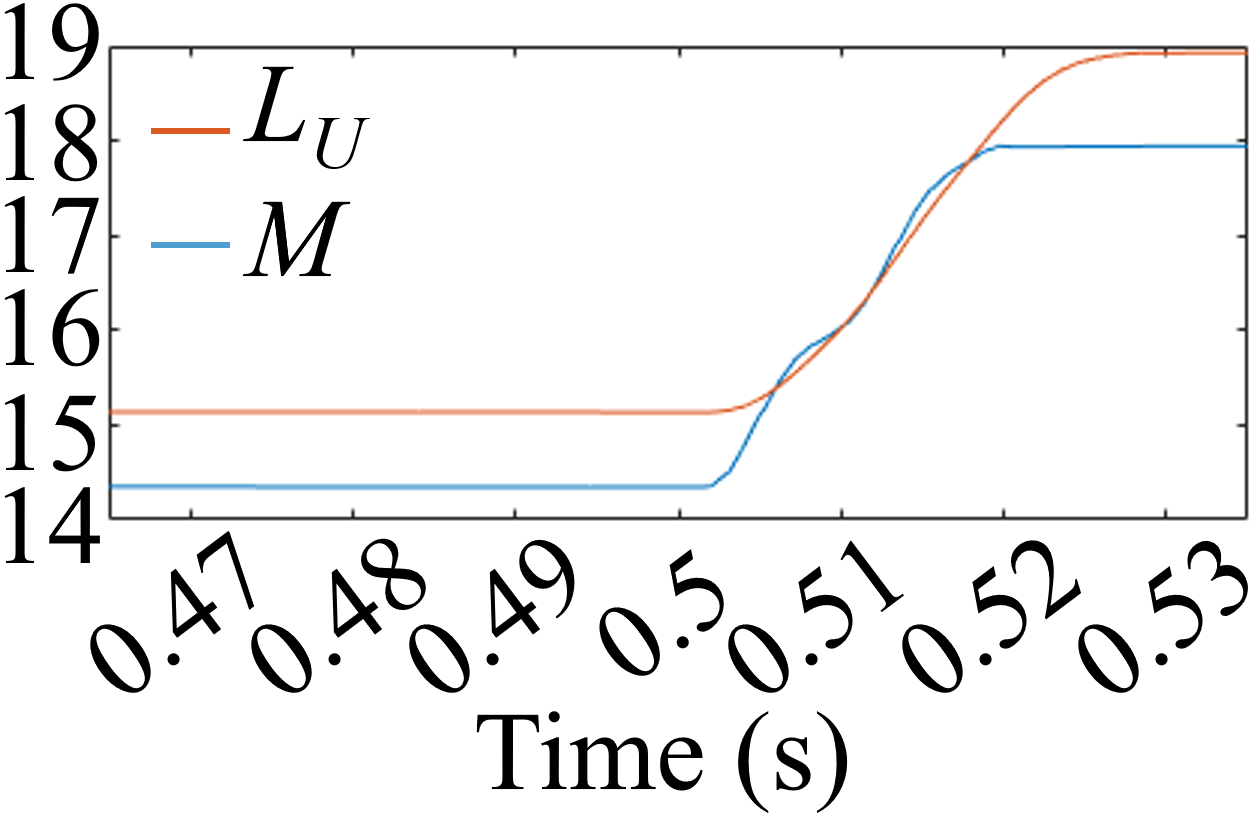}
\hspace{3pt}
\includegraphics[width=0.45\columnwidth]{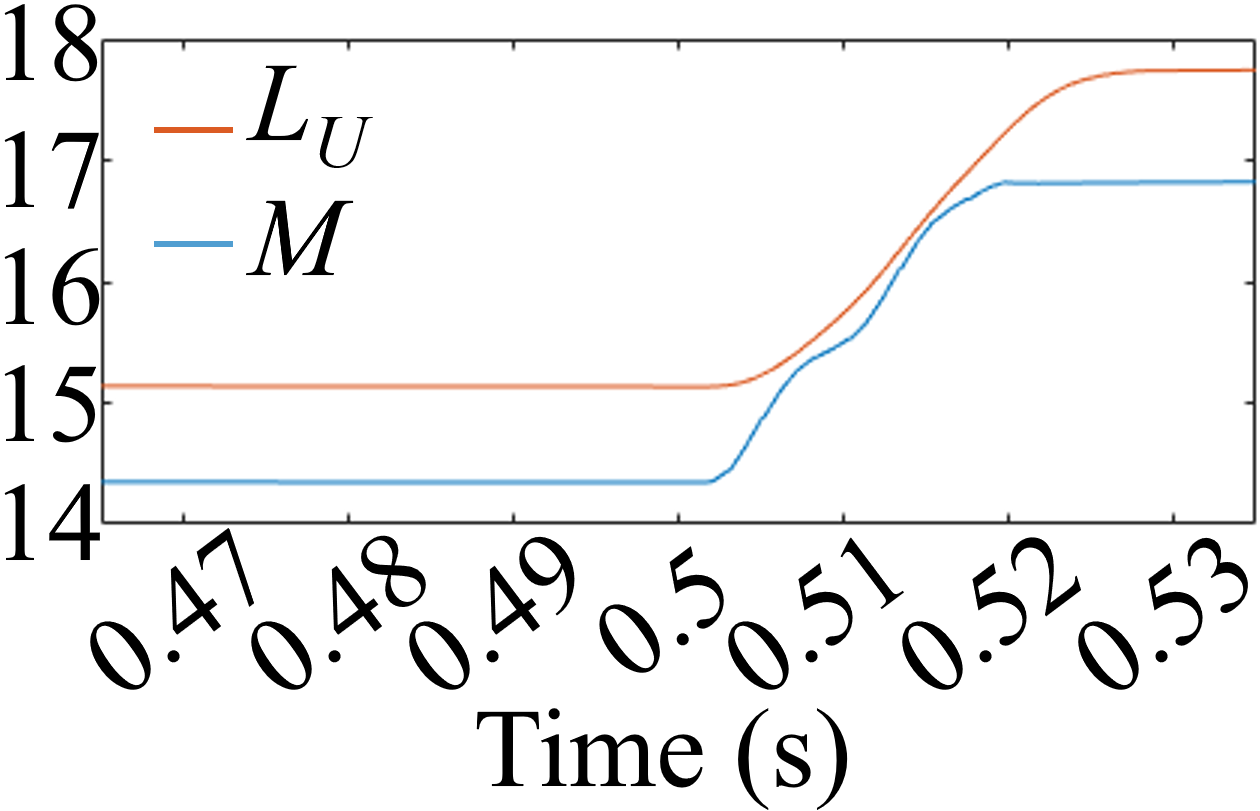}
\\
\hspace{4pt} (g) \hspace{102pt} (h)
\\
\vspace{5pt}
\includegraphics[width=0.45\columnwidth]{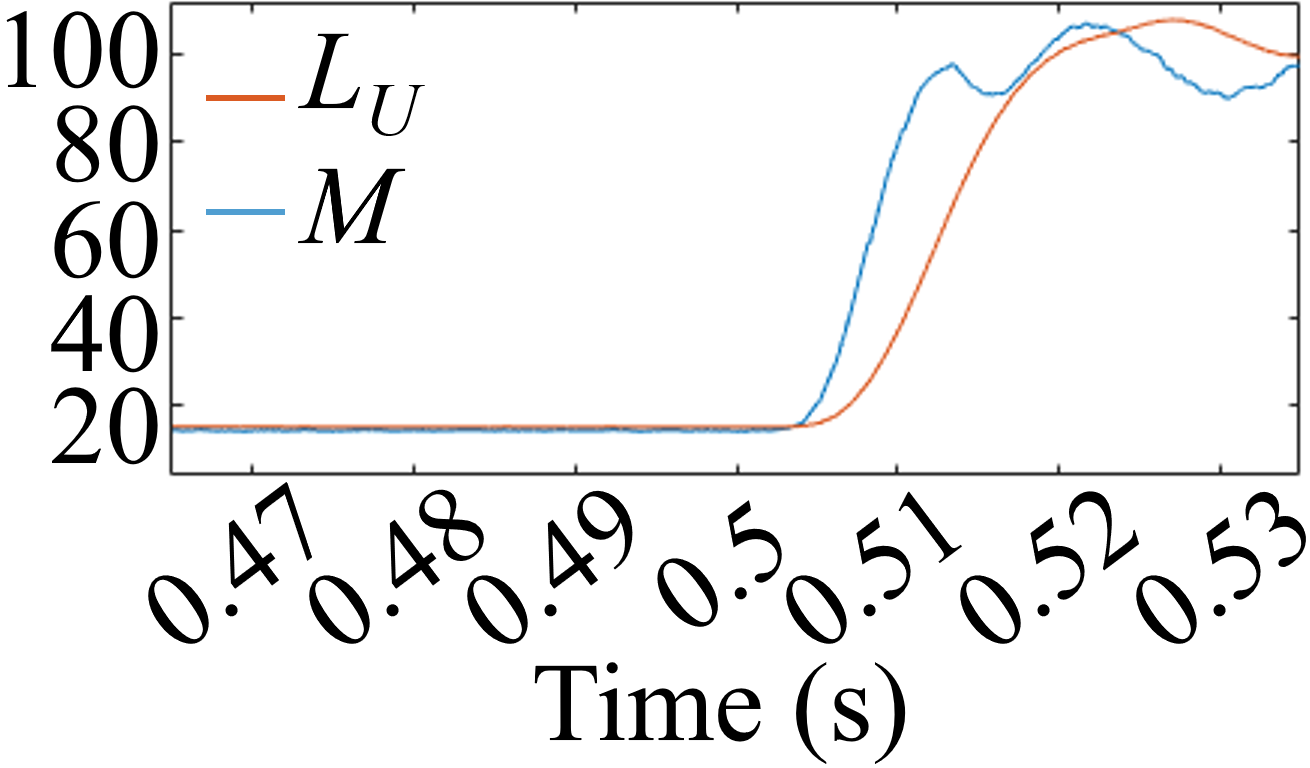}
\\ 
(i)  
\caption{Performance of the Mismatch Index under FMAs masking different faults. Illustrated masked faults are: 
(a)  \emph{A-G, bolted, $x$=10\%, } 
(b) \emph{A-G, bolted, $x$=50\%,  }
(c) \emph{A-G, bolted, $x$=90\%,}  
(d) \emph{C-G, bolted, $x$=10\%,  }
(e)\emph{ B-C, bolted, $x$=10\%,}  
(f) \emph{A-B-C-G, bolted, $x$=10\%,  }
(g) \emph{A-B-G, $R_f = 300 \Omega$, $x$=80\%,  }
(h) \emph{A-B-G, $R_f = 300 \Omega$, $x$=90\%, }
(i) \emph{A-G, bolted, $x$=10\%, with 35dB SNR.}
}
\label{fig:performance}
\end{figure}

Fig. \ref{fig:performance} showcases the performance of the proposed MI in detecting FMAs under different parameters. Firstly, three FMA cases masking single-phase faults, the most common fault type on transmission lines, are simulated.  Figures \ref{fig:performance}  (a), \ref{fig:performance}  (b) and \ref{fig:performance}  (c) show the MI under FMAs for bolted A-G faults at  $x$= 10\%, $x$= 50\% and $x$= 90\%, respectively, where $x$ is the percentage distance between the fault location and $LCDR_1$. Comparing Figures \ref{fig:performance}  (a)$-$\ref{fig:performance}  (c) reflects that close faults are easier for the MI to detect since they result in higher disruption of LCDRs' measurements. Next, Figures \ref{fig:performance}  (d)$-$\ref{fig:performance}  (e) illustrate FMAs masking a C-G, a B-C, and an A-B-C-G fault, respectively. These two cases share the same $R_f$ and fault location as the case in Fig. \ref{fig:performance} (a).
By comparing Figures \ref{fig:performance}  (d)$-$\ref{fig:performance}  (e) with Fig. \ref{fig:performance} (a), it can be noticed that the behavior of $M$ changes along with the fault type. Nonetheless, in all four cases, $M$ exceeds $L_U$, and FMAs are detected.
In addition, the effect of increasing the  $R_f$ is investigated. 
Figures \ref{fig:performance}  (g) and \ref{fig:performance}  (h) illustrate two FMAs masking 300-$\Omega$ A-B-G faults located at $x$=80\% and 90\%, respectively.
As the figures show, the proposed MI detects the FMAs with $300\Omega$ $R_f$ as far as 80\% of the line. Yet, at $x$=90\%, $M$ does not exceed $L_U$ for the same FMA. 
However, this value of $R_f$ is too high to achieve in practice. 
Moreover, 
FMAs masking faults of other types, e.g., symmetrical faults, are detected at the same high values of $R_f$ and $x$.
Finally, the effect of measurement noise is illustrated in Fig. \ref{fig:performance} (i), whereas the case in Fig. \ref{fig:performance} (a) is repeated but with 35 dB SNR. Comparing the two Figures reveals that measurement noise has a minor effect on the MI due to the moving averages in $M$ and $L_U$.

The first two columns of Table \ref{table:combined_results} summarize the obtained results for the MI only in the main case studies.
It can be noticed that the $TP$ and $FP$ percentages are  99.72 \%  and 14.55  \%, respectively. 
That is, 99.72 \% of FMAs are correctly detected, reflecting the high accuracy of the  MI. As a side effect, the proposed index flags $14.55$\% of outside-zone faults as possible cyberattacks. This side effect is dealt with using the second proposed module, as explained next.

\subsubsection{Performance Results of the Complete Proposed Solution} 

The proposed framework -comprising the MI and the trained ZCC- is then tested against the entirety of the two testing datasets for FMAs and external disturbances.
Using equations (19)$-$(25), simulation results confirm that, in this case study, the accuracy of the proposed solution is 99.845 \%, with a precision of 100\%, and a recall of 99.69\% as summarized in Table \ref{table:combined_results}.
On the one hand, no false alarms are observed, while, on the other hand, the percentage of successfully detected FMAs is 99.69\%. 
The small percentage of undetected FMAs are those cyberattacks masking unsymmetrical very high impedance faults located close to the far end of the protected line, which less frequently occur on large power systems. 
All the successfully detected FMAs are flagged within 1.5 power cycles, the time frame within which a commercial LCDR detects faults under no cyberattack.
These results reflect the proposed solution's capability to accurately and promptly detect FMAs with different fault attributes while maintaining a low false alarm rate.

\begin{table}[t]
\centering
\begingroup
\caption{Performance Results of The Proposed Solution}
\begin{tabular}{c|c|c}\hline
\centering
\makebox{Metric}
&\makebox{Mismatch Index only }
&\makebox{Mismatch Index + ZCC}
\\ \hline
\emph{accuracy}
& 92.58 \%
& \textcolor{black}{99.845 \%} \\
\emph{precision}
& 87.26 \%
& \textcolor{black}{100.00 \%} \\
\emph{recall}
& 99.71 \%
& \textcolor{black}{99.69 \%} \\
\textcolor{black}{\emph{TP} }
& \textcolor{black}{99.72 \%}
& \textcolor{black}{99.69 \%} \\
\textcolor{black}{\emph{TN} }
& \textcolor{black}{85.45 \%}
& \textcolor{black}{100.00 \%} \\
\hline
\end{tabular}
\label{table:combined_results}
\endgroup
\end{table}

Further, we evaluate the effect of varying the proposed solution's tuning parameters. We determine the Areas Under the receiver operator Curves (AUCs) for the MI only and the complete proposed solution. The curves are obtained by varying  $f$ of the MI, from 0.1\% to 10\%, and the ZCC's hyperparameters. Utilizing the MI only is found to have an AUC of about 0.929, compared to 0.99 in the case of the complete solution (MI + ZCC). These results reflect the robustness of the proposed solution under different tuning parameters. The smaller AUC associated with the MI is only because small values of $f$ increase the percentage of detected FMAs, leading to higher false alarm rates and vice versa.  However, this problem is tackled by the ZCC, which minimizes the $FP$ rate, boosting the proposed solution's accuracy as depicted in Table \ref{table:combined_results}.

\section{Sensitivity Analyses}

\subsection{Effect of Different System Loading Levels}

\begin{figure}[t!]
\centering
\includegraphics[width=1\columnwidth]{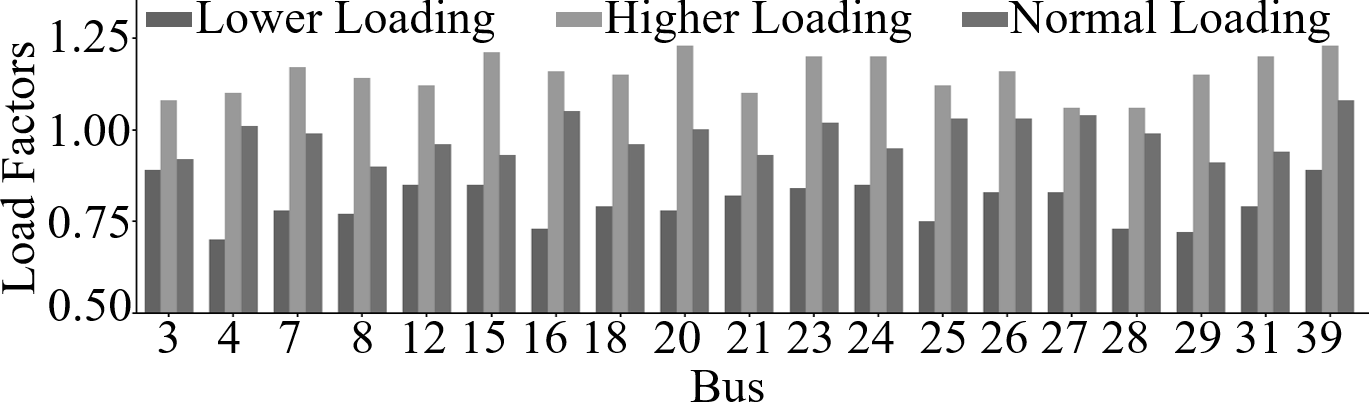}
\caption{Load factors utilized for different loading profiles.}
\label{fig:lf}
\end{figure}

Herein, the impact of the system's loading level is evaluated twice. For existing loads, the active and reactive powers are increased once and decreased the other time. This is performed by multiplying each load's original active and reactive powers by the load factors illustrated in Fig. \ref{fig:lf} \cite{Ahmadpaper}. For each of the two new system loading scenarios, 500 FMAs and 500 external disturbances are simulated, similar to the procedure explained in Section IV. The proposed framework is then tested against the newly generated cases. The  \emph{accuracy} is $99.34$\%, indicating the robustness of the proposed framework to variations in the system's loading level.

\subsection{Effects of Instrument Devices' Non-linearities}

Under severe faults that cause a significant voltage change, the Capacitive Voltage Transformer (CVT) that measures $V_1$ could face some transients. CVT transients increase the error and non-linearity in the voltage measurements, as the output of the CVT may not follow its input. The effect of CVT transients on the performance of the proposed framework is investigated. A total of 1000 FMAs are simulated on the line, with faults that cause depressed voltages, including low-resistance faults of varying types and locations. The proposed framework could successfully detect  99.8\% of these FMAs, reflecting its robustness to CVT transients.

\begin{figure}[t!]
\centering
\includegraphics[width=0.51\columnwidth]
{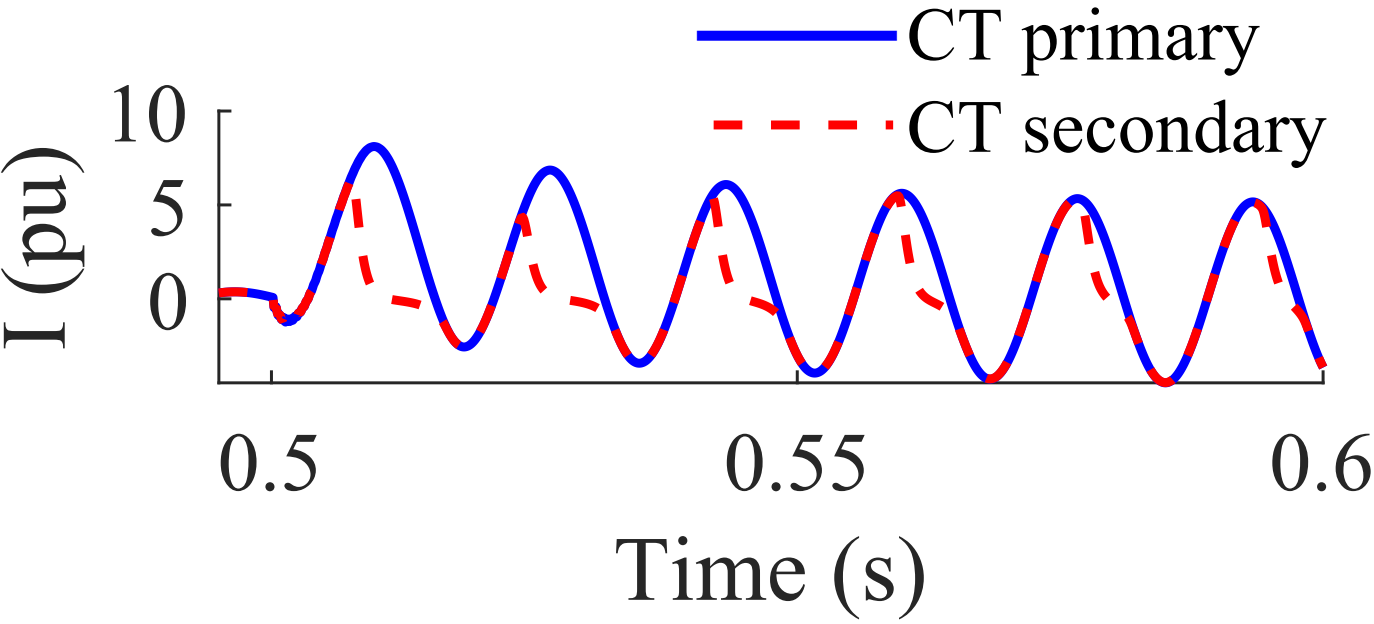}
\includegraphics[width=0.47\columnwidth]
{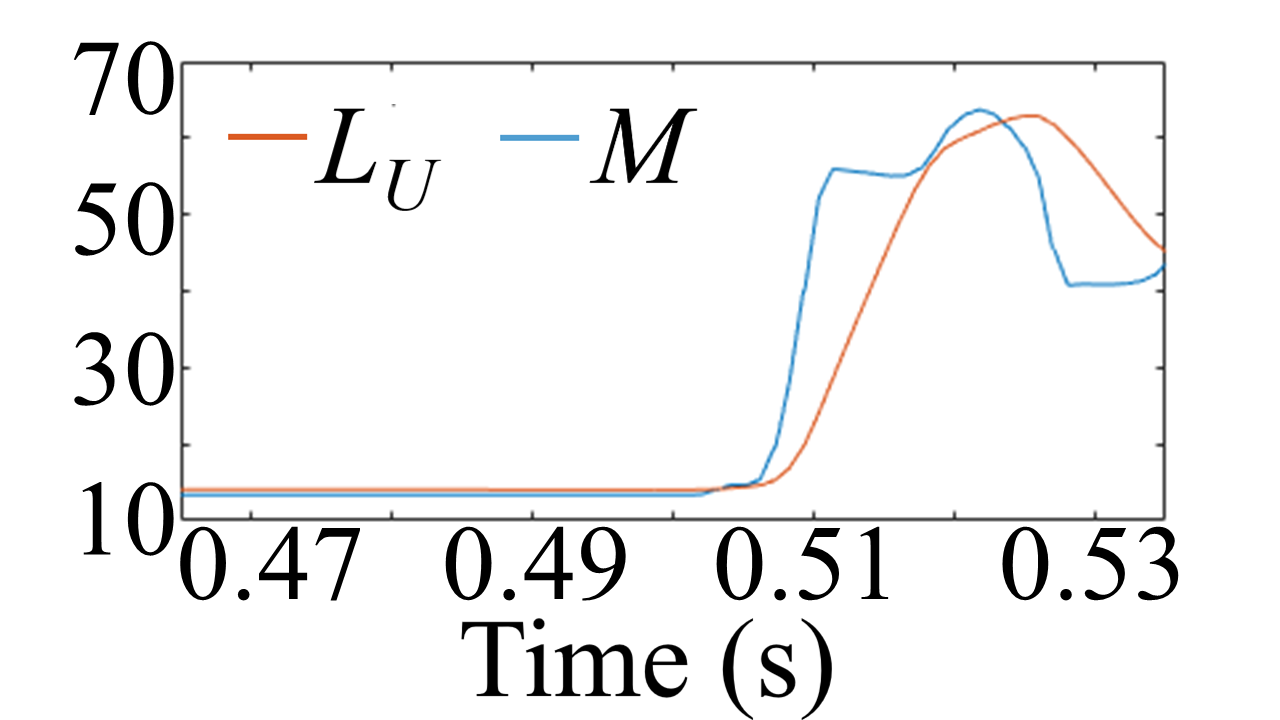}
\\
\hspace{4pt} (a) \hspace{85pt} (b) \\
\caption{ CT saturation case,
(a) currents,
(b) mismatch index's performance.}
\label{fig:CTsat_case}
\end{figure}

Similarly, if the CT  feeding the LCDR is inadequately designed, it may saturate under high-current faults. In the case of CT saturation, the current phasors received by the LCDR and the proposed framework, measured as per-unit quantities, can be of leading phase angles and reduced magnitudes compared to the per-unit current phasors on the primary side of the CT. To evaluate the performance of the proposed framework under CT saturation, a fault-masking cyberattack is conducted to mask a high-current A-C internal fault located at 50\% of line 11-6 with an $R_f$ of 0.5 $\Omega$. The measured current (CT secondary) is illustrated in Fig. \ref{fig:CTsat_case} (a). The figure shows that the current on the secondary side of the CT becomes of a lower magnitude and a leading phase angle compared to that on the primary side. Nevertheless, the current distortion does not prevent the MI from detecting the attack, as illustrated in Fig. \ref{fig:CTsat_case} (b), which is followed by the ZCC's confirmation.

\subsection{Effects of Sudden Dynamics and Multiple Internal Faults}

\begin{figure}[t!]
\centering
\includegraphics[width=0.45\columnwidth]
{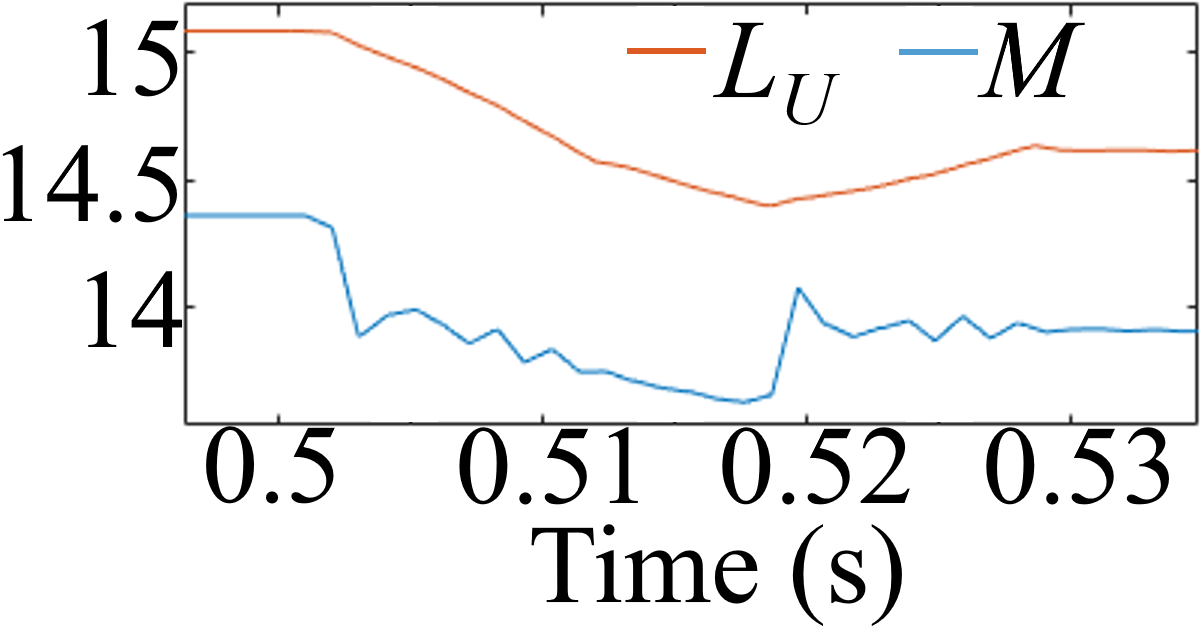}
\hspace{10pt}
\includegraphics[width=0.45\columnwidth]
{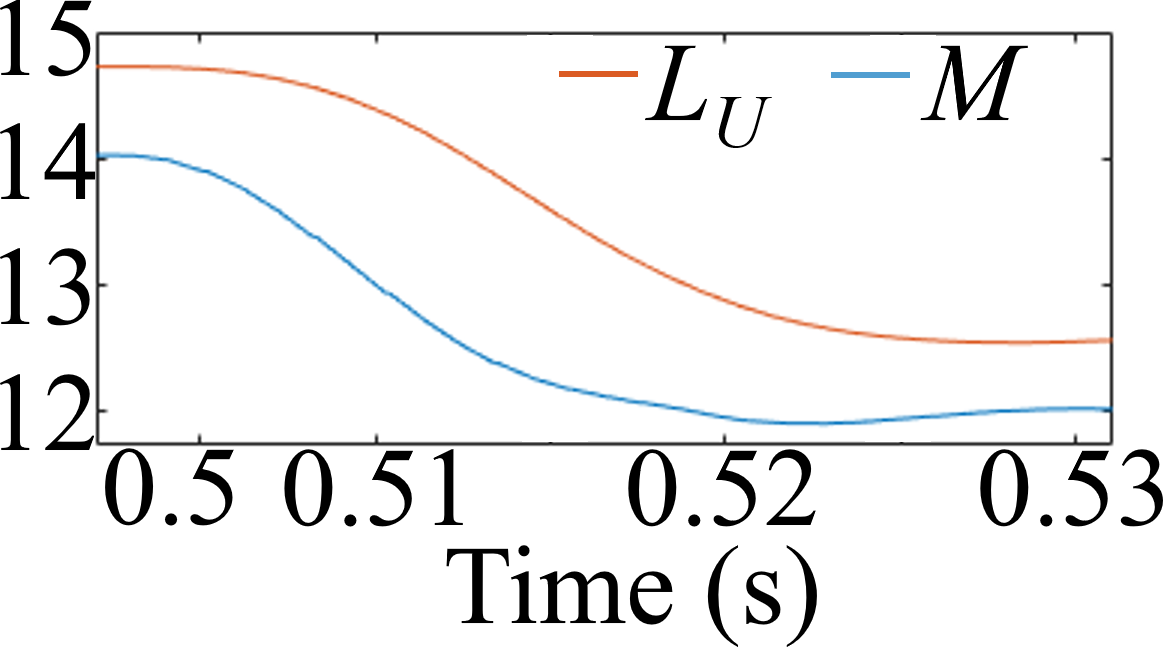} 
\\
\hspace{4pt} (a) \hspace{105pt} (b) 
\\
\vspace{5pt} 
\includegraphics[width=0.49\columnwidth]
{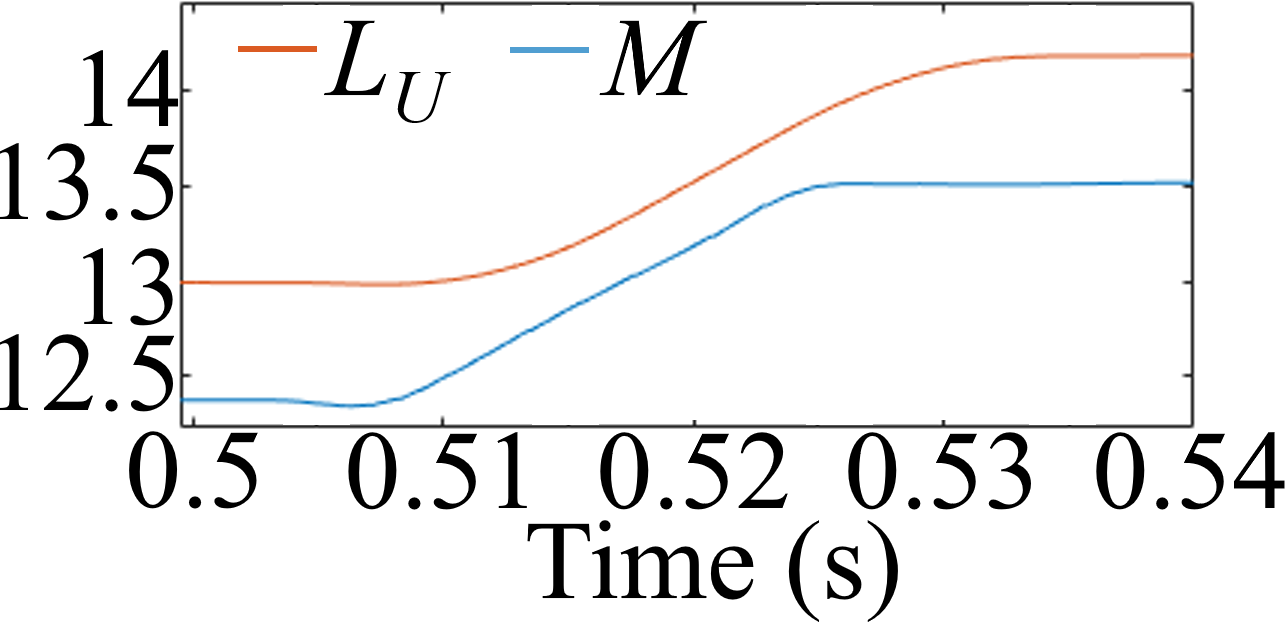}
\includegraphics[width=0.49\columnwidth]{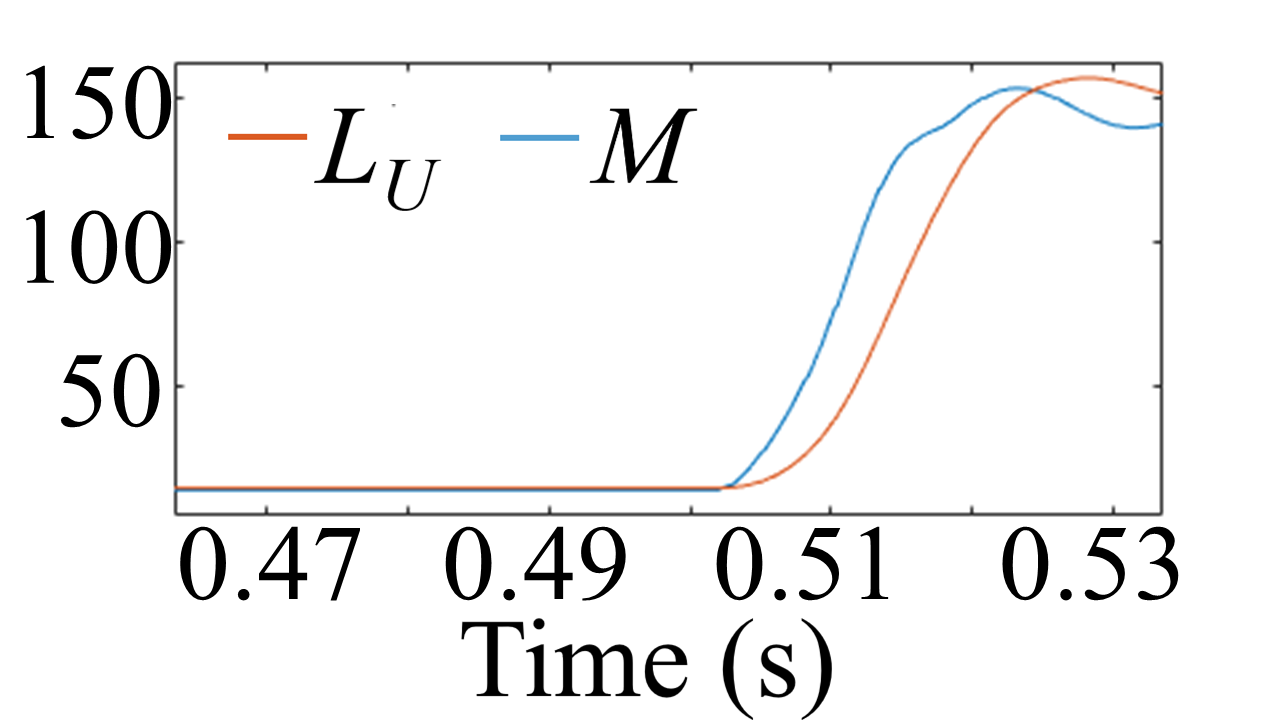}
\\
\hspace{4pt} (c) \hspace{105pt} (d)  
\caption{
Performance of the Mismatch Index under: (a) \Ablue{normal} capacitor bank switching, (b) \Ablue{normal} generation loss, (c) \Ablue{normal} generation connection, (d) \Ablue{an FMA masking simultaneous faults.}}
\label{fig:Sensetivity_Dynamics}
\end{figure}

Firstly, the impact of switching a nearby capacitor bank, installed on bus 6, is assessed. The 100 MVAR bank is switched on at 0.5 s, which is accompanied by a reactive power rush into line 11-6 and boosts the voltage at bus 6. 
Fig. \ref{fig:Sensetivity_Dynamics} (a) shows that this event perturbs the MI, as reflected by the sharp dynamics in $M$. However, the operating point of the MI remains below $L_U$ during the whole event. Next, the effects of generation loss and load loss are simulated. To perform these scenarios, the generator connected to bus 31 is switched off at one time and switched on at another, which is equivalent to a load loss. Our results confirm that the MI is not falsely triggered, as shown in Fig. \ref{fig:Sensetivity_Dynamics} (b) and Fig. \ref{fig:Sensetivity_Dynamics} (c).

Further, the performance of the proposed scheme is evaluated under an unusual but possible scenario.
A coordinated cyberattack is simulated to mask two simultaneous internal metallic faults: (i) a C-G fault with $x$= 10\%, and (ii) an A-B fault with $x$= 90\%. 
In general, simultaneous faults affect the magnitude and phase angles of fault currents, affecting all the measurements utilized by the proposed solution. 
Fig. \ref{fig:Sensetivity_Dynamics} (d)
illustrated how the proposed mismatch index is triggered by the fault-masking cyberattack in this case. 
Immediately afterward, the ZCC confirms that the triggering event is an FMA on $LCDR_1$, and the total detection time is a few milliseconds, i.e., is close to how fast $LCDR_1$ would normally detect the fault if there were no cyberattacks.

\subsection{Effect of Joint Cyberattacks}

\begin{figure}[t!]
\centering
\includegraphics[width=0.5\columnwidth]
{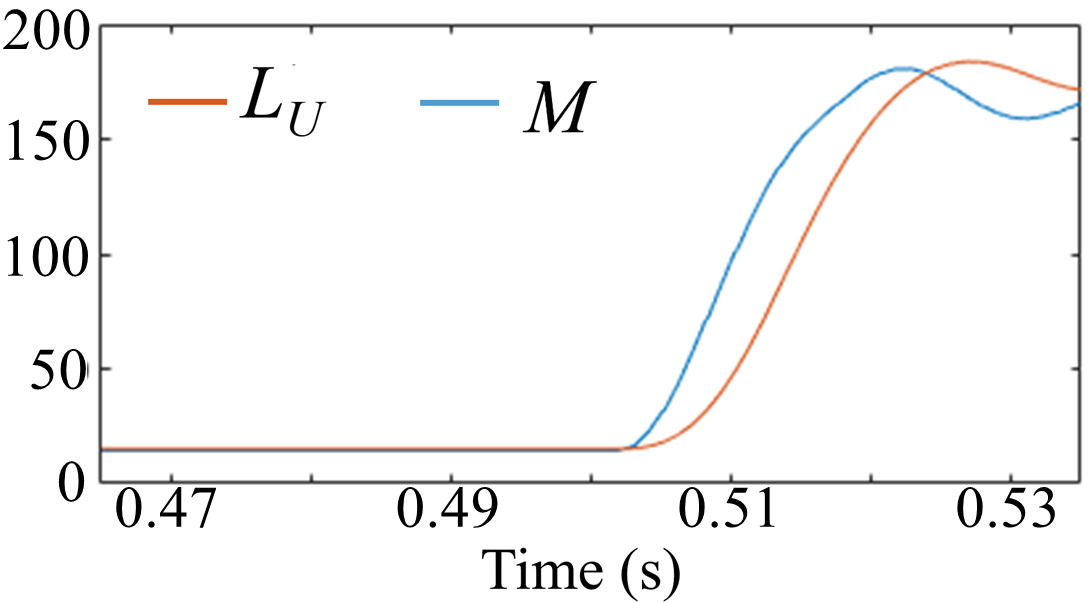}
\caption{
Performance of the Mismatch Index under joint attacks.
}
\label{fig:Sensetivity_Joint}
\end{figure}

Further, we evaluate the performance of the proposed two-stage FMA detection scheme under one of the worst cases where multiple LCDRS are jointly attacked. Two fault-masking cyberattacks are jointly carried out to mask two faults occurring at completely different parts of the power system. The occurrence of two distant faults affects the flow of the power and fault currents and, in turn, may result in the masked fault being undetected. 
In this scenario, a cyber FMA is simulated to mask a metallic symmetrical fault located at 50 \% of line 11-6, jointly with another FMA performed to mask a bolted C-G fault located at bus 29. 
Fig. \ref{fig:Sensetivity_Joint}
illustrates how the fault-masking cyberattack in this scenario triggers the proposed mismatch index. 
The ZCC also confirms that what triggered the mismatch index is an FMA on $LCDR_1$.

\section{ Comparative Analysis and Real-Time Verification }
Table \ref{table:Comparison_ours} depicts a comparative analysis between the proposed solution and existing methods. The proposed solution is fully dedicated to detecting FMAs on LCDRs.
Throughout our case studies, the proposed framework is shown to be more precise than the solution in \cite{Ahmadpaper_TII} (the only other work that considers FMAs on LCDRs) (100\% vs 99.01\% \textit{precision}). The higher \textit{precision} is achieved because the proposed two-stage framework focuses on detecting FMAs and also on minimizing the false positive rate by passing the received measurements firstly through the physics-based Mismatch Index followed by the ZCC neural network classifier, as discussed earlier. Additionally, the classifier module used in \cite{Ahmadpaper_TII} relies on the LCDR's remote measurements, making it prone to adversarial attacks, unlike the ZCC used in this work, which requires only the LCDR's local measurements.

\begin{table}[t!]
\centering
\begingroup
\caption{Comparative Analysis}
\begin{threeparttable}
\begin{tabular}{c | c | c  }\hline
\centering
\makebox{Solution}
&\makebox{Considers FMAs? }
&\makebox{Protection system}
\\   \hline 
\cite{Ahmadpaper, Kundur_DL, Remedial_pilot_main_protection, Amir1} 
& No & LCDR 
\\ 
\cite{Ahmadpaper_TII}\tnote{1} & Yes & LCDR 
\\  
\cite{Shahidehpour_Concealing, Shahidehpour_FMA} & Yes & WAP 
\\ 
This work\tnote{2}   & Yes & LCDR  
\\ \hline  
\end{tabular}
\begin{tablenotes}
    \item[1] False positive rate $\approx$ 1 \%.
    \item[2]  No reported false alarms.
\end{tablenotes}
\end{threeparttable}
\label{table:Comparison_ours}
\endgroup 
\end{table}

\begin{figure}[t!]
\centering
\includegraphics[width=1\columnwidth]{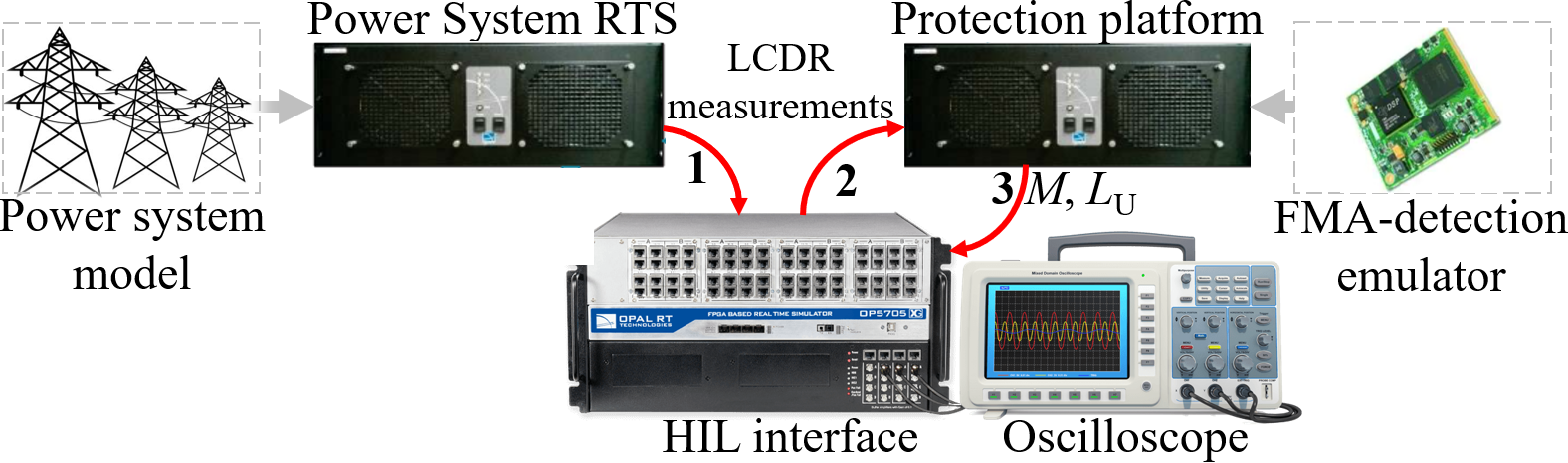}
\caption{Real-time test bed.}
\label{fig:HIL}
\end{figure}

Next, the performance of the proposed framework is verified using the real-time setup, illustrated in Fig. \ref{fig:HIL}, which comprises (i) an OP5700 RCP/Hardware-In-the-Loop FPGA-based Real-Time Simulator (RTS) \cite{OPAL}, (ii) an oscilloscope, and (iii) a computer for human interaction. In the RTS, quad-core processors run at 3.2GHz and utilize a 20M cache. A control/protection platform emulates the proposed FMA detection scheme within the RTS. This platform runs on one core of the RTS and is connected in a loop with another core that simulates the power system. In this environment, all simulation signals are generated in real time. 
To visualize the signals generated by the proposed framework, i.e., $M$, $L_U$, and the FMA alarm signal, the oscilloscope is connected to the RTS via its analog i/o ports.

\begin{figure}[t!]
\centering
\includegraphics[width=0.493\columnwidth]{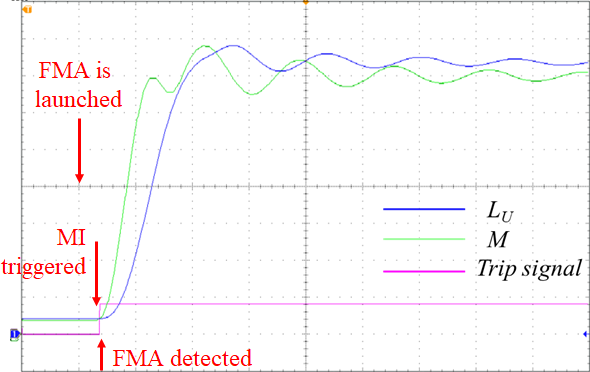}
\includegraphics[width=0.493\columnwidth]{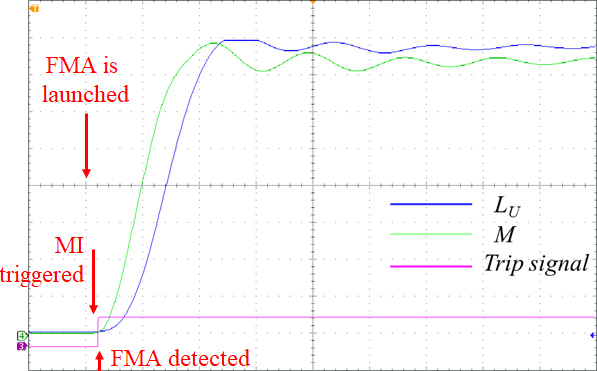}
\\
 (a) \textcolor{white}{............................................} (b) \\
 \vspace{9pt}
\includegraphics[width=0.493\columnwidth]{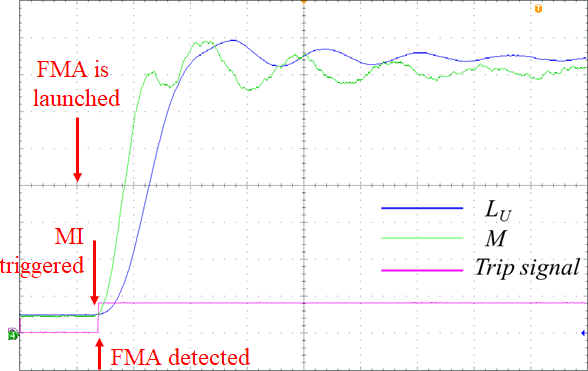}
\includegraphics[width=0.493\columnwidth]{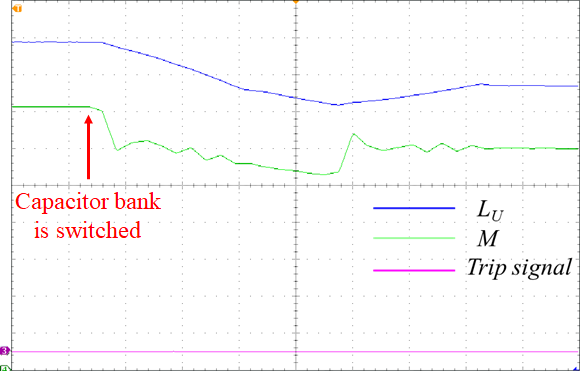}
\\
 (c) \textcolor{white}{............................................}  (d) 
\caption{Performance curves of the proposed framework captured from the oscilloscope, (a) FMA  (A-G solid fault, $x$=10), (b) FMA (A-B-C-G solid fault, $x$=10), (c) FMA (A-G solid fault, $x$=10, SNR= 35 dB), (d) cap.  switching.}
\label{fig:RTS_output}
\end{figure}

Using this setup, four scenarios from the previous sections, involving three cyber-induced FMAs masking faults of different parameters and a normal disturbance, are reproduced for real-time verification, as illustrated in Fig.  \ref{fig:RTS_output}.
In this figure, the blue and green curves represent the $M$ and $L_U$ of the MI, respectively. Moreover, the purple curve represents the FMA alarm signal issued by the proposed framework if the MI and the ANN are both triggered.
For visualization, $M$ and $L_U$ quantities are scaled down by a factor of 1/25 in sub-Figures (a)$-$(c), with a dc offset of -9. In sub-figure (d), these quantities are scaled up by a factor of 2.5 with a DC offset of -11.5. 
In the first scenario, Fig. \ref{fig:RTS_output} (a), an FMA masking a single phase fault similar to that in Fig. \ref{fig:performance} is simulated. Comparing both figures shows that the proposed framework can detect this FMA in real time. 
In the second scenario, illustrated by Fig.  \ref{fig:RTS_output} (b), a three-phase fault, the most severe type of fault, is masked by an FMA. However, the MI could successfully flag this FMA, and the ANN confirmed this decision, similar to the performance illustrated in the corresponding scenario in Fig. \ref{fig:performance} (f). The case of measurement noise is simulated in Fig.  \ref{fig:RTS_output} (c), where an FMA masks a single phase fault is recreated with the same attributes as the FMA in Fig. \ref{fig:performance} (i), i.e., in the presence of noise with 35 dB SNR. Herein, the proposed framework successfully detects this FMA in real time, similar to offline simulations.
In the three above FMA scenarios, the proposed framework detects the cyberattack within 5 milliseconds (the time scale of both oscilloscope channels is 10 ms).
Finally, the case of switching a nearby capacitor bank, previously illustrated in Fig. \ref{fig:Sensetivity_Dynamics} (c), is repeated in Fig.  \ref{fig:RTS_output} (d). Comparing these two Figures shows that the proposed framework, in real-time, avoids maloperation for non-FMA system dynamics such as nearby capacitor switching.

\section{Discussion}

\Ablue{While our framework performs well in detecting cyber FMAs used to mask faults of varying types, locations, and impedances, even under different power system operating points and measurement inaccuracies, attackers continuously adapt their strategies to evade detection. To address this challenge, it would be beneficial to explore strategies designed to detect and mitigate evolving threats. For example, machine learning techniques can effectively manage the challenge of adaptive attackers through frequent re-training \cite{shen2023increauth}, continuous model monitoring and updating \cite{continuous},  online learning \cite{online_learning}, and adversarial training \cite{adversarial}.}
Moreover, it will be interesting to investigate the potential of using explainable AI techniques compared to the black-box ANNs module. Developing physics-only-based protective relays that can detect both faults and cyber FMAs is another direction of research.

\section{Conclusion}

\Ablue{FMAs pose a serious threat to the safety and security of smart grids by stealthily concealing physical faults through cyberattacks.
This paper introduced a novel two-stage framework specifically designed to detect these covert cyber-induced FMAs on LCDRs.  }
The first module of the proposed framework, the line-model-based MI, detects mismatches in the LCDR's local and remote measurements that are likely to result from a masked fault. If the MI is triggered, the second module, an ANN-based ZCC, uses features extracted from the local measurements to confirm that the triggering event is a masked fault on the protected line before declaring an FMA. The proposed framework's performance was tested using the IEEE 39-bus benchmark system in PSCAD/EMTDC in various cases. The test cases included both FMAs and external disturbances under different system dynamics, operating conditions, and measurement errors. Our results confirm that the proposed framework (i) can accurately detect FMAs on LCDRs, (ii) does not maloperate for external faults and system disturbances, and (iii) is robust to variations in the system's loading level, measurement noise, CVT transients, and CT saturation. Further, the proposed framework's capability to operate in real-time is verified using OPAL-RT's RTS, \Ablue{which underscores the framework's practical applicability in modern power grids, offering a robust line of defense against cyber threats. Finally, future work directions have been discussed. }

\end{document}